\documentclass[fleqn,10pt]{wlscirep}
\usepackage[utf8]{inputenc}
\usepackage[T1]{fontenc}
\usepackage{siunitx}
\title{Simultaneous Bright- and Dark-Field X-ray Microscopy at X-ray Free Electron Lasers}

\usepackage{xcolor, soul}
\usepackage{upgreek}
\usepackage{array}
\newcommand{\um}{{$\upmu$}m }

\sethlcolor{yellow}

\author[1,2,3*]{Leora E. Dresselhaus-Marais}
\author[1]{Bernard Kozioziemski}
\author[4]{Theodor S. Holstad}
\author[4]{Trygve Magnus R\ae der}
\author[3]{Matthew Seaberg}
\author[5]{Daewoong Nam}
\author[5b]{Sangsoo Kim}
\author[6]{Sean Breckling}
\author[7]{Sungwook Choi}
\author[3]{Matthieu Chollet}
\author[8,T]{Philip K. Cook}
\author[1]{Eric Folsom}
\author[3]{Eric Galtier}
\author[6]{Arnulfo Gonzalez}
\author[9]{Tais Gorkhover}
\author[3]{Serge Guillet}
\author[4]{Kristoffer Haldrup}
\author[6]{Marylesa Howard}
\author[2,10]{Kento Katagiri}
\author[5b]{Seonghan Kim}
\author[5b]{Sunam Kim}
\author[7]{Sungwon Kim}
\author[7]{Hyunjung Kim}
\author[4]{Erik Bergb\"ack Knudsen}
\author[2,3]{Stephan Kuschel}
\author[3]{Hae-Ja Lee}
\author[12]{Chuanlong Lin}
\author[13]{R. Stewart McWilliams}
\author[3]{Bob Nagler}
\author[10]{Norimasa Ozaki}
\author[2,3]{Dayeeta Pal}
\author[14]{Ricardo Pablo Pedro}
\author[4]{Martin Meedom Nielsen}
\author[1]{Alison M. Saunders}
\author[15]{Frank Schoofs}
\author[12]{Toshimori Sekine}
\author[4]{Hugh Simons}
\author[3]{Tim van Driel}
\author[12]{Bihan Wang}
\author[12]{Wenge Yang}
\author[16,T]{Can Yildirim}
\author[4]{Henning Friis Poulsen}
\author[1]{Jon H. Eggert}
\affil[1]{Lawrence Livermore National Laboratory, Physics Division, Livermore, CA, USA}
\affil[2]{Stanford University, Department of Materials Science \& Engineering, Stanford, CA, USA}
\affil[3]{SLAC National Accelerator Laboratory, Menlo Park, CA, USA}
\affil[4]{Technical University of Denmark, Department of Physics, Kgs. Lyngby, Denmark}
\affil[5]{Photon Science Center, Pohang University and Science and Technology, Pohang, Korea}
\affil[5b]{XFEL Beamline Department, Pohang Accelerator Laboratory, Pohang, Korea}
\affil[6]{Nevada National Security Site, Las Vegas, NV, USA}
\affil[7]{Sogang University, Department of Physics, Seoul, Korea}
\affil[8]{BOKU, Vienna, Austria}
\affil[9]{University of Hamburg, Hamburg, Germany}
\affil[10]{Osaka University, Graduate School of Engineering, Suita, Japan}
\affil[12]{HPSTAR, Department, Shanghai, China}
\affil[13]{School of Physics and Astronomy, University of Edinburgh, Edinburgh, UK}
\affil[14]{Massachusetts Institute of Technology, Department of Nuclear Science \& Engineering, Cambridge, MA, USA}
\affil[15]{UK Atomic Energy Authority, Culham Science Centre, Abingdon, OX14 3DB, United Kingdom}
\affil[16]{CEA, Department, Grenoble, France}

\affil[*]{leoradm@stanford.edu}
\affil[T]{Now at ESRF, Grenoble, France}

\begin{abstract}
The structures, strain fields, and defect distributions in solid materials underlie the mechanical and physical properties across numerous applications. Many modern microstructural microscopy tools characterize crystal grains, domains and defects required to map lattice distortions or deformation, but are limited to studies of the (near) surface. Generally speaking, such tools cannot probe the structural dynamics in a way that is representative of bulk behavior. Synchrotron X-ray diffraction based imaging has long mapped the deeply embedded structural elements, and with enhanced resolution, Dark Field X-ray Microscopy (DFXM) can now map those features with the requisite nm-resolution. However, these techniques still suffer from the required integration times due to limitations from the source and optics. This work extends DFXM to X-ray free electron lasers, showing how the $10^{12}$ photons per pulse available at these sources offer structural characterization down to 100 fs resolution (orders of magnitude faster than current synchrotron images). We introduce the XFEL DFXM setup with simultaneous bright field microscopy to probe density changes within the same volume. This work presents a comprehensive guide to the multi-modal ultrafast high-resolution X-ray microscope that we constructed and tested at two XFELs, and shows initial data demonstrating two timing strategies to study associated reversible or irreversible lattice dynamics.

\end{abstract}
\begin{document}

\flushbottom
\maketitle

\thispagestyle{empty}

\section{Introduction}

Across materials science — from dislocation junctions strengthening materials to interstitial defects fracturing batteries over many charge cycles — defects change how materials respond to their surroundings\cite{Bulatov2006, Zhao2010}. Point defects are routinely used to finely tune material properties \cite{tuller2011pointdefects}, and defects extending across many unit cells (mesoscale) can tune the properties and performance of thermal or electronic materials, among others \cite{Zepeda-Ruiz2017, armstrong2021dislocation}. For example, grain boundaries in bismuth selenide have been shown to create nanodomains that enhance their thermoelectric efficiency by orders of magnitude by decoupling the mean-free-paths of electrons and phonons \cite{Goldsmid2001}. Similarly, in metals grain boundaries and dislocation networks govern bulk properties such as strength and ductility \cite{Russell2005}. At this time, our understanding and control of mesoscale defects and domains in bulk materials is primarily limited by our ability to probe their dynamics in a manner that is representative of bulk properties \cite{kubin1996dislocation}. The multiscale defect or grain structure often encountered imply that sample thicknesses of tens or hundreds of micrometers are required for representative sampling. Electron microscopy, field-ion microscopy and atom probe tomography can resolve defect cores with atomic resolution. However, they are intrinsically near surface probes and they rely on long raster scans to generate 3D maps, during which  sample conditions must be fixed \cite{ross2019tem,smith2013studies}. Without \textit{in-situ} measurement tools that can resolve how mesoscopic defects with nanometer cores interact to form large 3D networks that evolve over hundreds of micrometers, our understanding of the dynamics has been limited to theory that is yet untested at the microscopic scale. 

The primary challenge in detecting the mesoscopic structure lies in the wide range of length- and time-scales that must be probed to fully interpret the system. Lattice defects are comprised of local disruptions in the crystal packing – either a truncated plane (dislocation) or missing/extra atom (vacancy, interstitial), or truncated domain of the crystal (grain boundary). While the cores of defects have ~\AA-nm lengthscales, the long-range distortions from them that span micrometers to millimeters map key interactions that alter the macroscopic properties \cite{yildirim2022,armstrong2021dislocation,Bacon}. When these defects interact, the velocity of the property-transforming events can span from ballistic dynamics (ps-ns) through cumulative degradation (months to years), spanning $>$15 decades of timescales. A measurement tool to spatially and temporally resolve the evolution of plasticity \textit{in-situ} and, specifically, the interactions between adjacent strain or defects, requires sub-nanosecond imaging with nm-resolution \cite{Abbey2013,Campbell2014}. 

X-rays have been demonstrated to have the necessary penetration power to access this regime. The information encoded in these X-ray images is dictated by the material attribute responsible for the light-matter interaction that produces the beam being imaged. 
Based on their contrast mechanisms, X-ray microscopes can largely be divided into three modalities: spectral imaging, monochromatic phase/amplitude imaging, and diffraction-contrast microscopy
\cite{Abbey2013}. Spectral imaging gives contrast that delineates the elemental composition and sometimes the oxidation state, 
by scanning a microscopy measurement across an absorption or emission band of the material.
Monochromatic X-ray microscopy at photon energies far from an absorption or emission peak have contrast mechanisms described by the amplitude (e.g. radiography) or phase (e.g. ptychography) of the light that has traversed the material. Those techniques offer information about the density, thickness, and porosity of the material.
Finally, diffraction contrast microscopes map heterogeneity in the crystallography of the sample. While the contrast mechanism defines the materials-specific information contained in X-ray images, the resolution vs field of view are set by the imaging optics.

 While the contrast mechanism defines the materials-specific information contained in X-ray images, the resolution vs field of view are set by the imaging optics. Scanning microscopy is performed by focusing the X-ray beam to a small spot, then rastering the spot across the sample to collect spatially-mapped signals; for this approach, the resolution is set by the beam’s spot size while the field of view is set by the number of points in the scan. Tomographic microscopy requires a different approach to raster scans, collecting images from different rotational perspectives for a roughly cylindrical sample volume; the corresponding image stacks are compiled and Radon transformed to form the full 3D imaged volume. By contrast, full-field imaging collects information about the entire sampling volume in a single acquisition, with projections along one of the three sample dimensions. Near-field imaging (e.g. radiography, topography) captures full-field images very close to the sample, with a field of view limited by the detector or beam size and a resolution set by the Fresnel number of the image features. Magnified imaging with focusing optics that map the object onto its image plane in the far-field. Images collected along the X-ray transmitted beam (transmission X-ray microscopy, TXM) are collected along the bright-field (BF), and map the attenuation of the beam from absorption (and XRD in special cases). Conversely, images collected along the diffracted beam are produced by the X-rays scattering off of the undisrupted periodicity native to a given lattice plane, thereby mapping the crystallographic information along a specific symmetry of a crystal. These types of \textit{dark-field} (DF) images result in spatial maps of the disruptions to crystalline order in the lattice (i.e. the defects).


So called Dark-Field X-ray Microscopy (DFXM) is a full-field magnified microscope that acquires images by placing an X-ray objective lens along the X-ray diffracted beam. DFXM captures crystallographic distortions, i.e. the strain and mosaicity of the lattice, with a spatial resolution of 30-150 nm, with strain resolution of $10^{-5}$ and mosaicity (orientation) resolution of $10^{-3}$ radians. Recent work has used DFXM to characterize deformation texture \cite{yildirim2022snake}, dislocation structures \cite{Jakobsen2019}, domain boundary migration in ferroelectrics \cite{Simons2018}, fatigue in polycrystalline metals \cite{Sangid2020}, among other phenomena. We recently established time-resolved DFXM at the European Synchrotron Radiation Facility (ESRF) and demonstrated its utility with a first-ever study of collective dislocation dynamics deep inside bulk aluminum from $97-99\%$ of the melting temperature \cite{DresselhausMarais2021}. Today, DFXM has a  temporal resolution that is limited by the integration time required to acquire the images with synchrotron radiation;  access to material dynamics at sub-microsecond timescales requires more brilliant X-ray sources. 

The time resolution of X-ray microscopy tools is dictated by the acquisition scheme used for the measurement. For reversible and reproducible dynamics, a pump-probe modality may be used, whereby a ``pump'' excites a material into a transient state, then probe at some time-delay afterwards interrogates the relaxation of that state over fs-ps timescales, via millions of successive excitations that span all pump-probe delays and appropriate signal averaging. By contrast, irreversible processes require single-shot acquisitions with sufficiently high frame rates and signal to noise that they may gather all the relevant information about the system in the time that follows a single excitation pulse. Full-field imaging approaches are required to study irreversible processes.

Since the development of high photon energy (a.k.a. ``hard'') X-ray free electron lasers (XFELs), ultrafast X-ray science has had breakthroughs across physics, engineering, biology and beyond as optical and synchrotron techniques have become achievable with single-shot acquisition capabilities \cite{Yabashi2017}. With $10^{12}$ photons per pulse, XFELs have shifted static techniques that were hindered by their integration times to now become possible via ultrafast pump-probe measurements using femtosecond single-shot measurements could overcome issues of collection time and cumulative radiation damage \cite{Chapman2010}. For structural imaging, holography and X-ray coherent diffractive imaging (XCDI) have demonstrated this type of spatiotemporal resolution at XFEL sources \cite{Gorhover2016,Robinson2009}, but the necessary apertures and foci at the sample have prevented these methods from capturing a sufficiently large field-of-view to capture statistical populations of nanoscale features \cite{Sakdinawat2010}. While Bragg-XCDI has been able to spatially resolve the strain fields in nanoparticles \cite{Kim2018nanoparticle}, the same lattice-resolution measurements have not been extended to real-space X-ray imaging. XFEL imaging in real-space (i.e. in $x,y,z$ coordinates as opposed to XCDI's Fourier transformed raw images) has only been performed in the transmitted beam with associated density contrast \cite{barbato2022phasex}. This makes it insensitive to the sparse defects or anomalous strain fields that initiate plastic transformations or other macroscopic processes. To statistically probe the distortions that initiate large-scale material transformations, we need a technique with high spatiotemporal resolution and a large field-of-view that is sensitive to both density variation and localized strain fields inside the crystals. 

To address this challenge, in this work, we introduce DFXM at XFELs and present a multi-modal setup that enable simultaneous density mapping using BFXM (in the TXM geometry).  Using 32-fs X-ray pulses, local structural information is probed 13 orders of magnitude faster than previously accessible \cite{DresselhausMarais2021}. From experiments first at the the Pohang Accelerator Lab (PAL, 2019) and then at the Linac Coherent Light Source (LCLS, 2021), we present the instrumentation we have developed to build, align, acquire data, and analyze the results from the XFEL version of DFXM. To enhance the information-content of each frame and enable single-shot acquisitions of irreversible systems, we include simultaneous bright- and dark-field X-ray microscopy (BFXM \& DFXM), as they afford complementary information about the strain states and defect populations in sample. In these experiments, we benchmark our measurements using diamond single crystals. We first present our instrument design, including important trade-offs that must be considered when designing XFEL-DFXM experiments. We then present the analysis approaches we used for our data analysis, and present the types of data that may be collected at each of our 4 detectors in the imaging setup. We conclude with a discussion about the future opportunities for this instrument at XFELs around the world, across many fields of science.

\section{Dark-Field X-ray Microscopy Design}

\begin{figure}[!ht]
    \centering
    \includegraphics[width=0.95\textwidth]{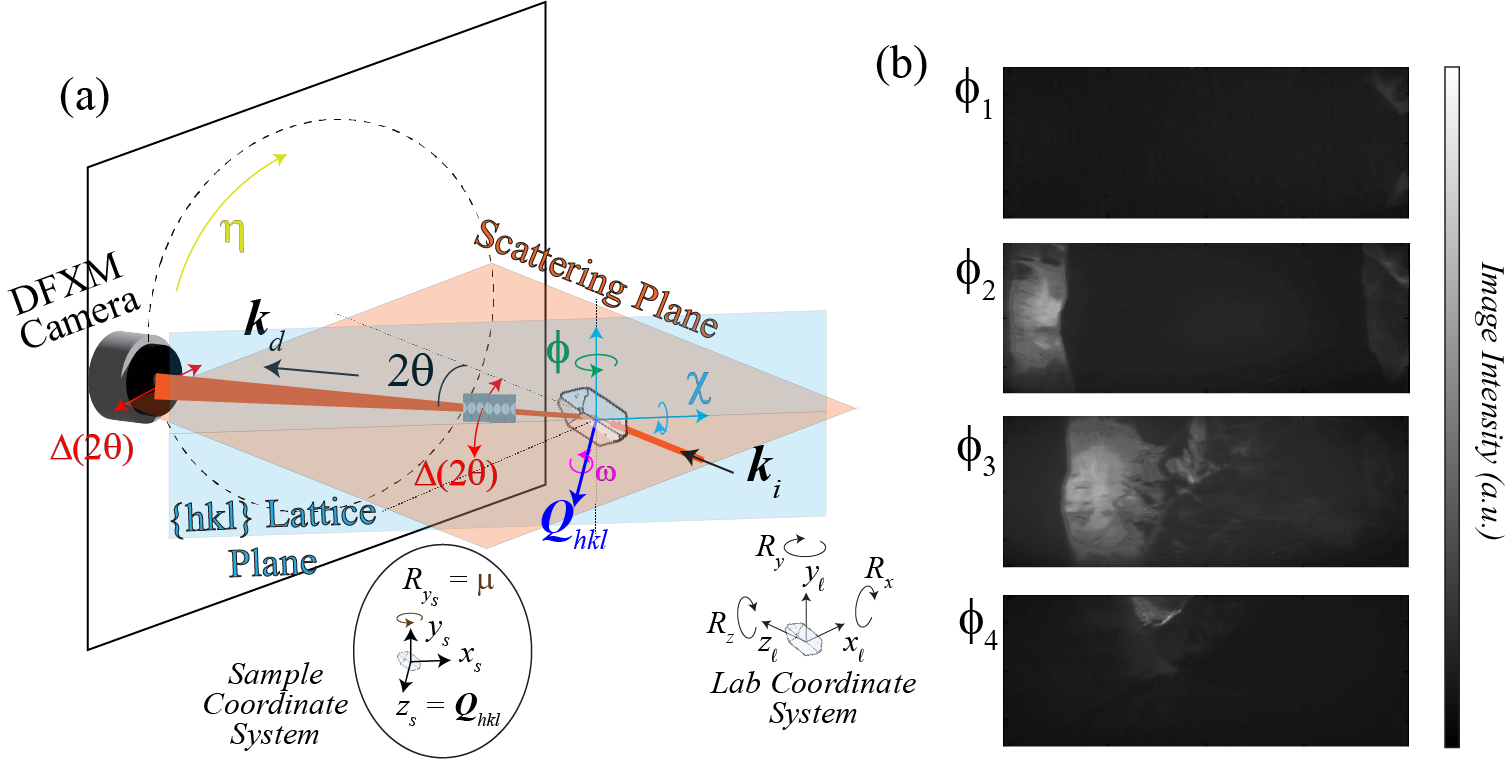}
    \caption{(a) Schematic showing the DFXM geometry and scanning procedure, including the rotational axes, $\chi$ (cyan), $\phi$ (green), and $\omega$ (magenta), the axial strain axis, $2\Delta\theta$ (red), the azimuthal detector position, $\eta$ (yellow), and the alignment rotation axis, $\mu$ (brown). The circular plot at the bottom illustrates how the $\mu$ motor is required to rotate all scanning stages from their initial orientations, $R_{x_{\ell}}$, $R_{y_{\ell}}$, and $R_{z_{\ell}}$ in the laboratory coordinate system, into the rotated sample coordinate system, \{$x_s,y_s,z_s$\} for which the diffraction vector $\vec{Q}_{hkl}$ are aligned to the $\hat{z}_s$ axis. (b) We also show images from representative points along a rocking-curve, showing intensity indicating the spatial populations of each point along the lattice distortion field for the same spatial plane in the sample.  Images in this stack were collected from single-crystal aluminum DFXM at the ESRF synchrotron at ID06-HXM.}
    \label{fig:Scanning}
\end{figure}

The DFXM geometry and the real- and reciprocal-space coordinate systems as used at synchrotrons are defined in Refs. \cite{Poulsen2017, Poulsen2018}. In this section, we introduce the design, coordinate systems, and axes for scanning, noting the key differences between the established synchrotron coordinate systems and the XFEL one for clarity.
We use the same notation here for consistency, with one main exception related to the direct space: the XFEL is defined to follow the optical conventions that defines the $\hat{z}_{\ell}$ axis as the beam's propagation direction (in the laboratory coordinate system). Given the complexity of the DFXM experiments, we detail a conversion from the synchrotron to XFEL systems here. For full understanding of how these coordinate systems convert to intensity and contrast mechanisms, see our previous work \cite{Poulsen2017,Poulsen2021,Holstad2022}. The full microscope setup in this orientation is shown in Figure~\ref{fig:Scanning}, with diffraction in the horizontal scattering geometry used at XFELs.

The analytical form to convert an arbitrary vector $\vec{v}$ from the laboratory coordinate system used in Ref. \cite{Poulsen2017}) into the XFEL one described in this work is
\begin{equation}
    \vec{v}_{\ell_{\text{XFEL}}} = \begin{pmatrix}
               0 & 0 & 1  \\
               -1 & 0 & 0  \\
               0 & 1 & 0
          \end{pmatrix} \vec{v}_{\ell_{\text{synch}}}. \hspace{5 mm}
\end{equation}

As shown in Figure \ref{fig:Scanning}, the laboratory coordinate system is defined by $\{x_{\ell},y_{\ell},z_{\ell}\}$ and rotations about those axes, $\{R_{x_{\ell}},R_{y_{\ell}},R_{z_{\ell}}\}$, defined by the Poynting vector of the incident beam, $\vec{k}_i$. When the beam reaches the sample, it interacts with the lattice planes in the relevant grain based on the orientation of the lattice. We have previously defined a crystal coordinate system to describe how the crystallographic vectors, $\vec{a}$, $\vec{b}$, and $\vec{c}$ map onto the lab system using $\{x_{c},y_{c},z_{c}\}$, such that the transform matrix $\bf{M}^{\ell \xrightarrow[]{} c}$ converts any arbitrary vector from the lab to crystal system \cite{Poulsen2021}. 

After traversing the sample, the diffracted beam propagates along a diffracted wavevector, $\vec{k}_d || \hat{z}_i$, that defines the diffraction imaging coordinate system, $\{x_{i},y_{i},z_{i}\}$, based on rotation about the $\hat{y}_{\ell}$-axis. 

With DFXM, subtle changes to the contrast mechanism are provided by minor angular offsets of the diffraction vector from the nominal value for the undeformed lattice $\vec{Q}_{hkl}$ based on crystal rotations or by varying the scattering angle $2\theta$ \cite{Poulsen2018}. The latter motion requires a coherent translation of the far-field (ff) detector and imaging lenses along the $2\theta$ arc. 
As can be seen in Fig.~\ref{fig:Scanning}, a combined $2\Delta\theta$-$\phi$ scan will probe the local axial strain. The local variation in crystal orientation, as expressed by the pole figure of the reflection (a.k.a. the mosaicity), is probed by varying the goniometer settings, corresponding to rotations $R_{y_{s}}$ and $R_{z_{s}}$.  These are known as rocking and rolling scan, respectively. 

The XFEL instrument constraints currently require diffraction in the horizontal scattering plane due to space, polarization, and infrastructure constraints, as shown by the orange plane in Fig. \ref{fig:Scanning}. Since $\vec{k}_i$ and $\vec{k}_d$ are defined in the horizontal plane, the $2\theta$ value and diffraction plane constrains the $\vec{Q}_{hkl}$ vector orientations that may be probed for crystals in this microscope. \textit{As shown by the blue plane in Fig. \ref{fig:Scanning}, the $\{\omega,\chi,\phi\}$ rotation vectors that define the goniometer are explicitly defined in reference to the direction of ${\vec{Q}}_{hkl}$, meaning that they define the relationship between the crystallographic and sample coordinate systems.} If the sample's surface normal was already cut with $Q_{hkl}$ aligned to the surface normal of the crystal, this describes a rotation of $\theta$ about the $\hat{y}_{\ell}$ axis in the lab system. We note that the sample coordinate system \{$x_s,y_s,z_s$\} describes the motorized axes with respect to the surface normal of the sample with respect to the goniometer, while a separate crystallographic coordinate system, \{$x_c,y_c,z_c$\}, allows one to account for the crystallographic orientation of that crystal with respect to its $uvw$ vectors (scaled inverse of $hkl$).
The components of $\mathbf{M}^{\ell \xrightarrow[]{} s}$ detail the conversion from the rotations about the $\{x_{\ell},y_{\ell},z_{\ell}\}$ system into the goniometer axes, $\{\omega,\chi,\phi\}$, as is inspired by Busing and Levy \cite{busing1967goniometer}. 

As in previous work, DFXM can collect image series' along scans that sample reciprocal space while rotating the crystal about $\omega$, $\chi$ or $\phi$; by translating the lens and detector along $2\theta$ and $\eta$; or by translating the sample along $z_s$. It is worth noting that, while only 2 rotation stages are technically required to access the full sphere of rotation, the precision of this experiment's high sensitivity necessitates rotation stages about 4 axes for accurate scanning. As described in Poulsen, et al.,\cite{Poulsen2017,Poulsen2021} a bottom rotation stage, $\mu$, is required to orient the 3-axis Eulerian cradle ($\omega,\chi,\phi$) to align the $\omega$ rotations to the ${\vec{Q}}$ vector. The vector defining the relationship between each coordinate system is given for the XFEL coordinate system in Table~\ref{tab:coordinate_systems}.
We note that in contrast to the previous work, $\chi$ does not rotate about the incident beam axis, $z_{\ell}$, but instead rotates about an axis that is rotated by an angle of $\theta$ with respect to $\vec{k}_i$, to ensure the rotations occur about the principal axes of reciprocal space. We note that $2\theta$ describes the scattering angle probed by the diffracting plane, while $2\Delta\theta$ describes the strain scanning axis required to probe the material via $\theta$-$2\theta$ scans.

\begin{table}[h]
\centering
\renewcommand{\arraystretch}{1.3}
\begin{tabular}{|c|c|c|c|}
    \hline
    \textbf{Lab System} & \textbf{Sample System} & \textbf{Crystal System} & \textbf{Imaging System} \\[0.5ex]
    \hline
    $\hat{k}_i = \hat{z}_{\ell}$ & $\hat{n}=\hat{z}_s$ & $\hat{Q}_{hkl}=\hat{z}_c$ & $\hat{k}_d=\hat{z}_i$ \\
    \hline
    \hline
\end{tabular}
\caption{Coordinate systems to describe DFXM with relevant defining characteristics, given in the XFEL coordinate system that differs from the one described in Poulsen et al.~\cite{Poulsen2021}.}
\label{tab:coordinate_systems}
\end{table}

\section{Experimental methods}
We detail in this section the experimental methods we used to implement DFXM at the PAL-XFEL and subsequently at the LCLS. This section details the full instrument design details to explain the intuition necessary to design the appropriate microscope for specific applications. We begin explaining the selection of beam parameters and upstream optics (Section \ref{subsection:upstream optics}), then proceed to explain an overall picture of the imaging optics and microscope specifications (Section \ref{subsection:microscope design}). We follow this with a detailed description of the goniometer stages and the rationale for our design choices in that based on the precision required for the sample and objective alignments (Section \ref{subsection:gonoimeter design}). We then detail our choice of imaging detectors and our procedure for the calibration of those detectors in Section \ref{subsection:detectors}. We close with a description of our alignment cameras and overall alignment strategy, with detail on the impact of these alignments in interpretability of the results (Section \ref{subsection:alignment}). 

\subsection{Upstream Beam-Conditioning Optics}
\label{subsection:upstream optics}
While the setup configurations differed slightly between experiments at the PAL-XFEL and the LCLS, both studied diamond single crystals to demonstrate proof of concept for the technique under different sample conditions. The PAL-XFEL experiment used a 9.7-keV X-ray pulse with $\sim 10^{11}$ photons per 32-fs duration pulse at a 30-Hz repetition rate; we prefocused the beam to a 30$\times$10-\um$^2$ spot at the sample. Our DFXM experiment at LCLS used a 10.1 keV X-ray pulse with 10$^{12}$ photons per 50-fs per pulse with a repetition rate of 30 Hz. The energy of each XFEL pulse was calibrated using the Intensity-Position Monitors (IPMs) installed into the XCS hutch, as described in \cite{alonso2015x, feng2011single}.
The LCLS beam was prefocused using a stack of 1D and 2D Be Compound Refractive Lenses (CRLs) with varied radii of curvature (focusing power), placed 3.435-m upstream of the sample to horizontally focus the beam into a vertical line beam at the sample position. From front to back, the 9-lenslets in the stack were: of two 2D lenses of $R = 100$-\um, two 2D lenses of $R = 500$-\um, one 1D lens of $R = 200$-\um, three 1D lenses of $R = 300$-\um, and one 1D lens of $R = 500$-\um. 
In both experiments, the photon flux was calibrated for each pulse and used to normalize the resulting images to correct for beam fluctuations. 
Since our first experiment at PAL-XFEL informed our full design for the refined DFXM microscope designed for XFELs, we focus our instrumentation descriptions on the setup in full for the latter experiment at LCLS.

At the LCLS, our experiment used a channel-cut Si monochromator to reduce the bandwidth to $\Delta E/E \sim 10^{-4}$, with a beam divergence of $1.1 \times 1.1 \mu$rad$^2$. 
A monochromatic beam with a stable spectrum is essential to be able to acquire interpretable results with DFXM because fluctuations in the incident beam's photon energy change the $d$-spacing and orientation that are imaged by DFXM \cite{Poulsen2021,Holstad2022}, as we discuss fully in Section~\ref{subsection:alignment}.
After passing through prefocusing lenses, the resulting beam was apertured further with power slits and then cleanup slits to reduce the size of the focused line beam to its minimum 3.4 $\times$ 3805-\um dimensions at the sample (as measured via curve fitting of BFXM images). To observe the beam positioning and pump-probe overlap, we installed a removable scintillator at the sample position and a viewing camera in reflection mode on the upstream side of the sample.

When selecting prefocusing lenses, it is important to achieve the narrowest beam waist at the sample position to ensure minimal integration over the observation plane that comprises the image. Placing condenser lenses with low focal length near the entry-surface of the crystal achieves the narrowest waist in the $x$-direction, but can shorten the depth of focus, making the Rayleigh range over which the beam is a constant size much smaller. In general, we have found that at XFELs, the longer prefocusing distances tended to be required due to space constraints, however, this does limit the interpretability of the results in some systems.

\subsection{Multi-Modal microscope design}
\label{subsection:microscope design}
\begin{figure}[h!]
    \centering
    \includegraphics[width=0.9\linewidth]{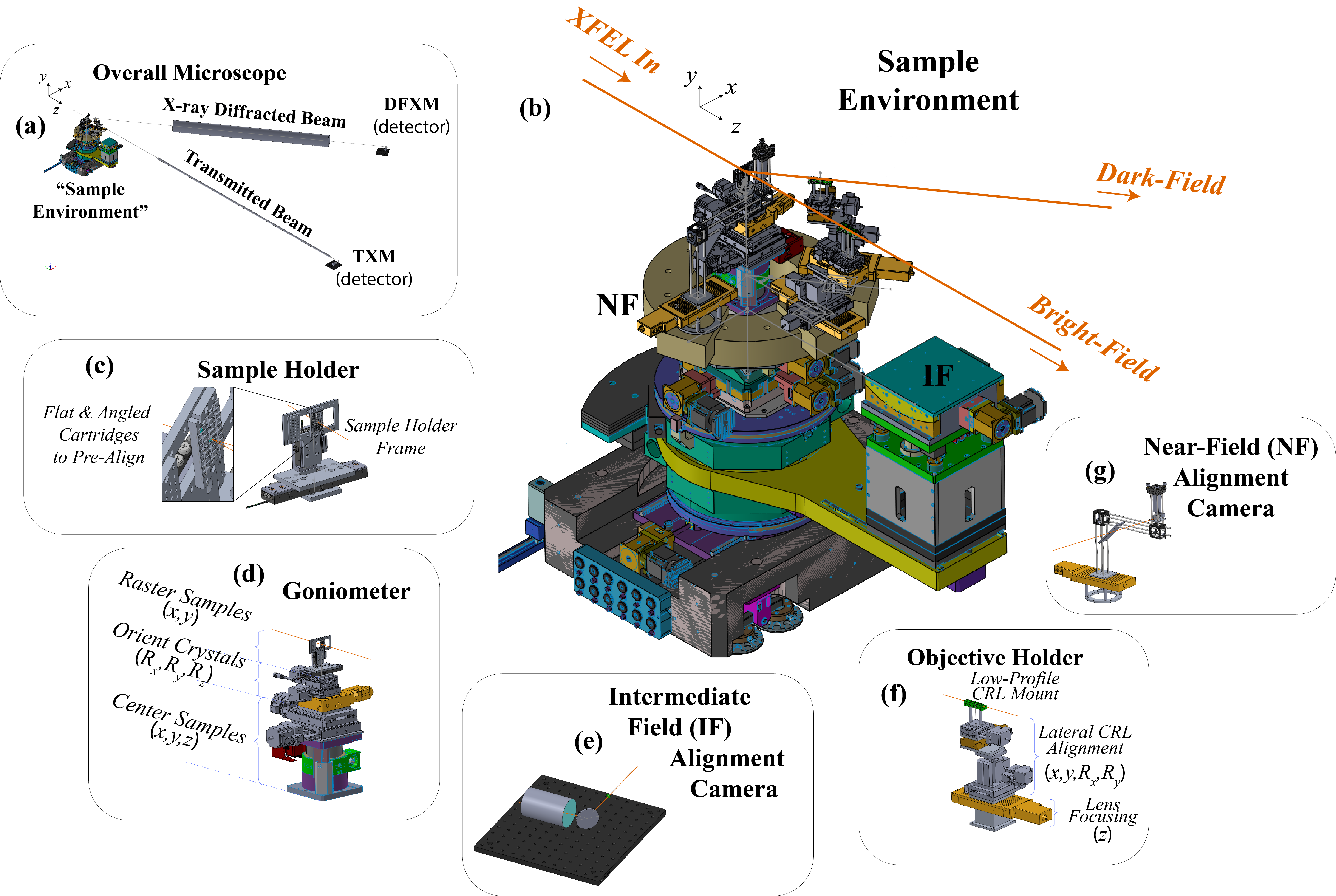}
    \caption{3D models of the different stage assembly at the sample position. (a) An overview of the entire setup, including the sample environment assembly, the beam tubes, and the far-field detectors, to show the scale of the experiment. (b) A zoomed in perspective of the Sample Environment, showing the arrangement of each camera, stage assembly, and imaging beam surrounding the sample position. The remaining images show zoomed-in isolated models of each components described in the text, including (c) the sample holder, (d) the goniometer assembly, (e) the intermediate field (IF) camera, (f) the CRL holder, (g) the near-field (NF) detector.}
    \label{fig:Setup}
\end{figure}

After the X-ray pulse traversed the Bragg-oriented sample, its transmitted and \{$111$\} diffraction beam was imaged onto two far-field (FF) detectors, one along the transmitted and one along the diffracted beam (Fig.~\ref{fig:Setup}). Along the transmitted (direct) beam, TXM was installed. The imaging objective was a stack of $N=17$ Be 2D CRLs (radius of curvature, $R=50$-\um, distance between lenslet centers, $T=2$-mm), corresponding to an effective focal length of 42 cm and a pupil aperture (FWHM) of 261 \um. The working distance (sample to entry of CRL) was set to $d_1=47$-cm while the distance from the exit- surface of the CRL to the detector was $d_2=8.437$-m  (total sample-to-detector distance $d_{\text{tot}}=8.941$-m). From this follows a theoretical magnification of the X-ray beam of $\mathcal{M} = $ 18.4$\times$.
The relevant formalism for calculating the optical parameters are given in \cite{Poulsen2017}. The stray unfocused light was blocked with $600$-\um diameter pinholes at the front and exit surfaces of the CRL stack, cut from 1-mm thick copper.

For the DFXM imaging system, the CRL comprised $N=33$ Be 2D CRLs ($R=50$-\um, $T=2$-mm), corresponding to an effective focal length of 20 cm and a pupil aperture (FWHM) of \SI{370}{\micro\meter}. This objective was aligned with the center of the \{111\} diffracted beam. With $d_1=23$-cm and  $d_2=6.532$-m (and total sample-to-detector distance $d_{\text{tot}}=6.832$-m), the calculated magnification is $\mathcal{M} = $ 27$\times$. As the larger $N$ stack for the DF imaging system produced a smaller effective aperture, we used smaller Cu pinholes of $300$-\um diameter (1-mm thick) at the front and back faces of the stack to remove stray unfocussed light. 

Long $d_2$ distances along both imaging systems attenuates the beams significantly if propagating through air. Attenuation was therefore mitigated using vacuum beam-tubes that spanned the longest possible extent of the $d_2$ beam path, though there was slight attenuation from the 125-\um thick kapton windows at each end. We placed a 50-\um thick sheet of aluminium foil at the front surface of our vacuum tube along the BFXM imaging arm to avoid saturating the detector with the unattenuated XFEL beam. In our case, the $2\theta = 35^{\circ}$ limited the need to account for the polarization factor in our microscope design ($P = 0.84$), however, the coherence of the XFEL beam requires consideration for the polarization factor to ensure optimal signal to noise in the experiment \cite{Ishikawa2005}.

Diamond single crystals cut with surface normals of $[110]$ were placed into the interaction point with the XFEL beam; the 660-\um thick crystals were oriented to an angle of 37.5$^{\circ}$ about the $R_y$ axis (i.e. rotated about $y_{\ell}$ from being normal to the incident X-ray beam) in the direction opposite of the diffracting beam to meet the Bragg condition for the $\{111\}$ plane. The 1D X-ray beam thus illuminated an angled but nearly planar sheet of 3.4$\times$3805$\times$1151-\um$^3$ volume in the sample, with the FWHM of the line focus of 3.4 \um. The region of interest (ROI) for each imaging system gave different views of the beam  along the bright-field (BF) and dark-field (DF) imaging arms because of the different projection vectors, causing the integrated volumes differ between experiments. DF's diffraction projects the wavefront along a new wavevector, $\vec{k}_{\text{d}}$ (i.e. deflected it by $2\theta$ in the horizontal plane). As such, the TXM images mapped the ($x_{\ell}, y_{\ell})$ plane of the sample, integrating along the  $z_{\ell}$-axis, while the DFXM images projected a map of the $(z_{\ell}, y_{\ell})$ plane, integrating the images along $x_{\ell}$ (the line-beam's narrowest waist). The numerical apertures, pinholes, and magnification of the two imaging systems made each one's field-of-view (FOV) differ slightly: TXM mapped 4$\times$101-\um$^2$, while DFXM mapped 700$\times$200-\um$^2$ (or less, depending on the diffraction condition).

The DFXM had a total magnification of 54$\times$ (corresponding to 0.11 \um per pixel), while the TXM along the bright-field had a total magnification of 14$\times$ (corresponding to 0.44 \um per pixel).
We note that the approximate magnification may be predicted based on the imaging equations and effective focal lengths described previously, however, since CRLs natively have aberrations, the actual magnification must be measured explicitly. At synchrotrons, this is often done either using resolution targets (e.g. Siemens stars, US Air Force Targets, TEM grids, etc.), or by moving translation motors to observe translation of the features on pixels. We note that at XFELs, the motorized approach is often highly inaccurate, as the precision, accuracy, and reproducibility of all motors are not always reliable down to the sub-\um lengthscales required by this experiment, and are subject to change based
on the setup at any given implementation of the setup. As such, we strongly encourage the use of resolution targets, and used TEM grids (200- and 1500-mesh) for this work.
Finally, we note the importance of understanding the orientation of the observation plane. As in X-ray topography (XRT), the contrast mechanism giving rise to image signal in DFXM arises from scattering along the diffraction vector of the lattice \{$Q_{hkl}$\}. That said, \textit{the spatial plane that is imaged by DFXM is always rotated by an angle of $\theta$ with respect to the observation plane (as shown in Fig.~}\ref{fig:pixelSize}b).

\subsection{Goniometer Design}
\label{subsection:gonoimeter design}

For quick alignment of this sensitive 2-path (TXM, DFXM) imaging experiment, we designed a custom goniometer to streamline the alignment and scanning precision. We defined our initial design criteria based on the points listed below:
\begin{itemize}[noitemsep,nolistsep]
    \item Fully self-contained goniometer assembly for accurate pre- alignment before the XFEL beam turned on.
    \item Precise \& reproducible positioning of sample translation ($x_s,y_s,z_s$) \& rotation ($\chi,\phi,\omega$) with sub-\um and 0.005$^{\circ}$ precision, respectively.
    \item Motor controls for fast translation between alignment targets (scintillators, resolution targets) and samples of interest (without manual realignment).
    \item Rapid orientation of each crystal into its optimized Bragg condition in the horizontal scattering plane.
    \item Scanning \& data acquisition capabilities to collect series' of images while scanning the sample \& detector along $x,z,\chi,\phi,\omega,2\theta$.
    \item Independent motion control for each CRL stack along its local $x_i,y_i,R_{x_i},R_{y_i}$ with respect to the sample position along each imaging beam.
    \item Viewing cameras near the Fourier plane of each imaging CRL stack for alignment (possible future aperturing opportunities).
    \item Facile motion of the CRLs along an arc concentric with the samples for direct-beam alignment of both stacks.
\end{itemize}

Based on the needs outlined above, we designed the goniometer and stage assemblies shown in Figure \ref{fig:Setup}(a-b). 
We used a 12-axis goniometer, using (from top to bottom): 2 piezo motor translation stages ($x,y$) to position each sample of interest into the center of rotation for the rotation stages, 3 rotation stages to mimic an Eulerian cradle with rotation about the $z$, $x$, and $y$ axes (i.e. $R_x$, $R_y$, $R_z$), 3 linear translation stages to center the pre-defined center of rotation into the center of rotation for the CRL carousel, and finally, a 2-axis linear translation stage ($x$,$z$) to position the goniometer's center of rotation into the XFEL interaction point.  

To simplify CRL alignment and strain scans, we used a motorized $2\theta$ rail that could independently rotate 3 carriages about its center, as shown in Fig~\ref{fig:Setup}. The radius of this circle was selected based on the imaging distances for 10 keV photon energies and $\sim$33 imaging CRLs, and the baseplate for each carriage was planned to integrate radial translation stages for further focusing of each lens stack. Each carriage was fitted with stage stacks to position the BF and DF stacks of CRLs along their respective ($x,y,R_x,R_y$) axes, as shown in Fig. \ref{fig:Setup}b. The third carousel was used for the near-field alignment camera, though in future this could be used for 2-phase DFXM (i.e. diffractive imaging for a second phase).

At the top of the goniometer, we required resolution targets, alignment guides, calibration standards and single crystals for our measurements. For facile motion between these, we used a sample holder with interchangeable custom cartridges that could affix onto a sample holder frame that was stationed at the image plane of the microscope; cartridges were designed to hold
targets either at an arbitrary geometry, or at a pre-aligned Bragg condition, depending on the intended use of the samples it intended to hold. 
The cartridges in our case were machined for high precision positioning, but in future we anticipate they could be 3D printed before or during an experiment for unexpected sample/ diffraction geometries.
Pre-positioning the samples in this way simplified the goniometer design and alignment needs because it limited the range of angular sweeps to $\lesssim$ $\pm 5^{\circ}$ required to orient the crystals.

\subsection{Imaging detectors}
\label{subsection:detectors}
Selecting the appropriate imaging detectors for this experiment requires consideration of the dynamic range, resolution, and frame rates. 
Indirect X-ray detection (i.e. conversion of X-ray to optical light via scintillator crystals) offers the highest spatial resolution, as the magnification may be amplified by a relay visible- light imaging system that offers higher fidelity/ magnification lenses and smaller pixel sizes on the visible-light cameras. However, this comes at the cost of a much lower photon detection efficiency.
The experiments in this work used home-built indirect detectors (similar to those employed in \cite{koch1998x}) for the far-field (FF) cameras along both TXM and DFXM arms, using 50-\um thick Ce:LuAG scintillator crystals (thicker crystals correspond to decreased spatial resolution). Each scintillator crystal was placed into the X-ray image plane, then was relay imaged with $\sim$1x magnification to form an optical image on the visible- light camera. We used an Optem 34-11-10 zoom-lens attached to an Andor Zyla 5.5 sCMOS camera for both detectors, with scintillator-lens distances of 285-mm along the TXM arm and 130-mm along the DFXM arm. This resulted in an optical magnification of 0.71$\times$ for the TXM arm and 0.75$\times$ optical magnification for the DFXM arm. A turning mirror was placed between the scintillator and zoom lens to avoid burning the detector and lens with unconverted X-ray. We used both cameras at full resolution, but over a limited ROI that spanned the spatial extent of the scintillator screen to achieve the detector readout rates of 30-Hz. Since the light under-filled the active area of each detector, this did not limit our field of view in the material.

For all detectors in this work, we calibrated the spatial resolution and magnification using TEM grids. For each detector, we placed a TEM grid (200-mesh) at the upstream surface of the scintillator crystal to calibrate the optical magnification based on comparisons to the pixel sizes of each camera. We then calibrated the magnification of each imaging system using an illuminated TEM grid placed at the sample position. As the sample holder (Fig.~\ref{fig:Setup}c) held all calibration and imaging samples at the same $z$-position defining the object, we were able to calibrate the BF imaging systems with a bare TEM grid held at normal incidence to the sample, and the DF imaging systems using a TEM grid affixed to the exit surface of a single-crystal of diamond. The exit surface was important to avoid effects from dynamical diffraction from hindering the measurement of the
diffraction image.

\begin{figure}[h]
    \centering
    \includegraphics[width=0.9\textwidth]{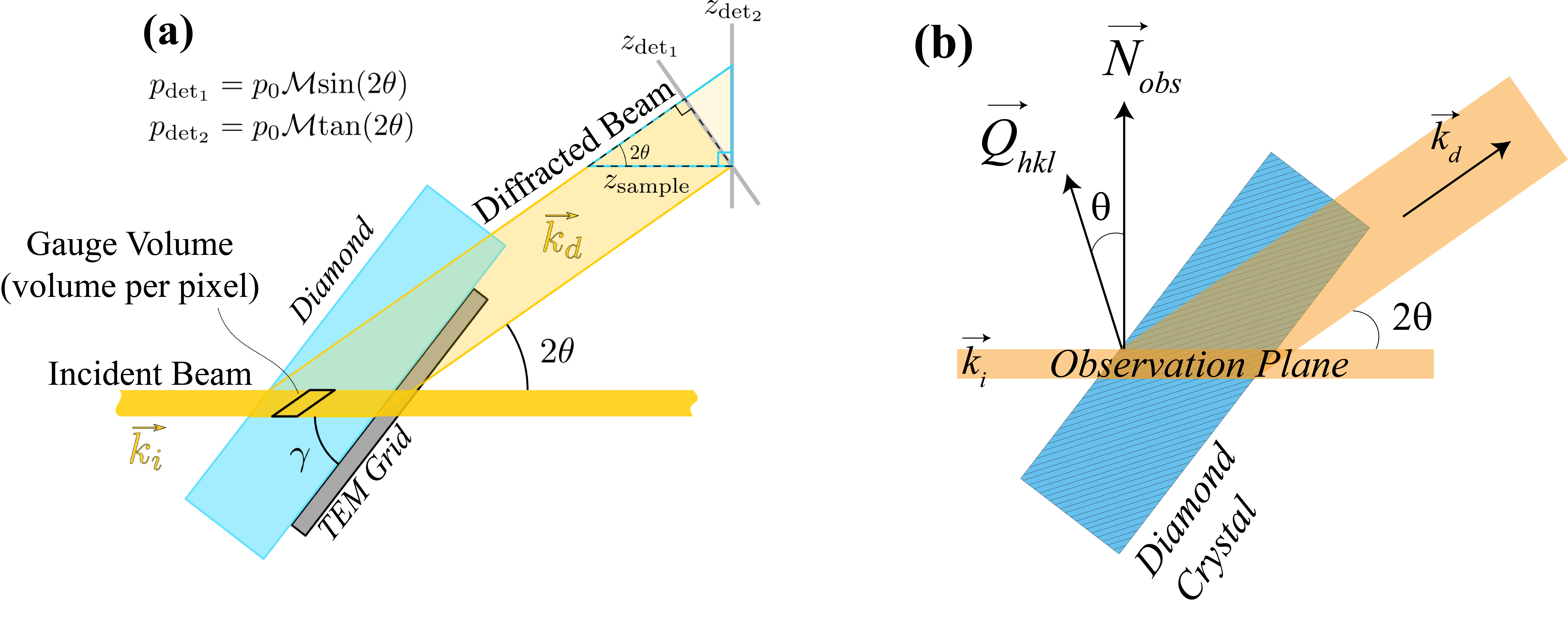}
    \caption{(a) Schematic showing a view from the top (along $y_{\ell}$) of the microscope that illustrates how a single pixel is projected onto the detector in two different detector orientations: $\hat{z}_{\text{det}_1}$ to describe the orientation with the detector face perpendicular to the incident beam ($\vec{k}_i \perp \hat{z}_{\text{det}_1}$) as done in \cite{Poulsen2021}, and $\hat{z}_{\text{det}_2}$ to describe the orientation with the detector face perpendicular to the diffracted beam ($\vec{k}_d \perp \hat{z}_{\text{det}_2}$). (b) Schematic showing the relationship between the observation plane normal, $\vec{N}_{obs}$, and the scattering vector, $\vec{Q}_{hkl}$} 
    \label{fig:pixelSize}
\end{figure}

Indeed beyond the image magnification, the processing of images collected with DFXM and XRT require image stretching to account for their projection angles. Because diffraction-contrast images are measured along the X-ray diffracted beam, they map the $(z,y)$ components of the sample by projecting the illuminated observation plate at the angle $2\theta$. While the line-beam simplifies image interpretation, it also necessitates careful calibration of image stretching to correct for this scattering projection, as shown in Figure~\ref{fig:pixelSize}.

Because the LADM detector is oriented at normal incidence to the diffracted XFEL beam, the stretch in the images differs from what has been described for imaging at ID06 at ESRF (shown in Figure~\ref{fig:pixelSize} as $z_{\text{det}_2}$) \cite{Poulsen2021}. If we consider that the detector orientation at the XFEL is perpendicular to the Bragg-scattered beam ($det_1$), and that the standard detector orientation at ID06 is normal to the direct beam ($det_2$), the appropriate expressions relating the pixel size on the detector $p_{det}$ to  the effective pixel size $p_0$ in the observation plane along $z_{\ell}$ is
\begin{equation}
    p_{\text{det}_1} = p_{0z} \cdot \mathcal{M} \cdot \sin(2\theta), \newline
    p_{\text{det}_2} = p_{0z} \cdot \mathcal{M} \cdot \tan(2\theta).
\end{equation}
as shown graphically in Figure \ref{fig:pixelSize}. The effective pixel size along the non stretched direction $y_{\ell}$, retains the equation, $p_{det} = p_{0y} \cdot \mathcal{M}$ in both cases. 
We note that the TEM grids affixed to the exit surface of the diamond crystals necessitate that the resolution calibrations be performed with an additional stretch of the image by a factor of $\cos{\gamma}$ to account for the orientation of the crystal (also annotated in Figure~\ref{fig:pixelSize}a).

Finally, we note the importance of understanding the orientation of the observation plane. As in XRT, the contrast mechanism giving rise to image signal in DFXM arises from scattering contrast along the \{$Q_{hkl}$\} diffraction vector of the lattice. That said, \textit{the spatial observation plane that gives rise to DFXM images is always rotated by an angle of $\theta$ about the scattering plane normal with respect to the reciprocal lattice vector of the crystal's diffracting plane.} The crystallographic notation in legends for all acquired images shown in this work are all offset by the $\theta$ rotation for simplicity, but careful analyses of DFXM images requires calibration for the observation-plane's orientational offset for accurate interpretation.

\subsection{Alignment strategy}
\label{subsection:alignment}

As each imaging system had a long beam path, we installed alignment cameras at three additional viewing positions: (1) upstream of the sample, (2) immediately behind the sample, in the near-field (NF), and (3) slightly downstream of the CRLs, in the  ``intermediate field'' (IF). The upstream camera was primarily for alignment of the sample positioning stages - to confirm pump-probe overlap, XFEL positioning, and sample motion. The two alignment cameras downstream of the sample confirmed the orientation of the crystals for alignment to the far-field detectors. As the IF camera was located behind the imaging CRLs, it also served as an alignment guide for the imaging CRLs. 

\textbf{Viewing Camera.} The first alignment camera was a viewing camera, placed upstream of the sample holder, that was a simple optical camera with a Zoom lens that resolved the interaction point of the XFEL with our sample. This camera allowed us to effectively position the sample and identify the position of the XFEL on our holder mount. 

\textbf{Near-Field (NF) Camera.} On the third carriage of the $2\theta$ carousel stage, we placed a NF camera 6.5-cm behind the sample, along the direct beam. We then positioned it using a $\pm100$-mm linear translation stage, to identify the position of the diffracted beam, which appeared at the appropriate angle of 34.75$^{\circ}$ (with respect to the $0^{\circ}$ reference along the transmitted beam). In this way, our NF detector could capture images along the BF and DF beams, allowing us to optimize the crystal's orientation into the horizontal Bragg condition, which is quite strict with the coherent and monochromatic XFEL beam \cite{Holstad2022}. This NF camera also enabled us to orient the CRL stacks to the appropriate angles around the carousel stage, and to optimize the $R_z$ alignment ($\sim$ $\omega$) to orient the crystal's $\vec{Q}$ such that the \{111\} beam scattered into the horizontal plane. This low magnification detector used a 35\um-thick Ce:LuAG scintillator crystal to convert from X-ray to optical light, then relay imaged the visible light with a Zoom lens onto an Allied Vision Mako g-319b camera. Its 3.45-\um pixels produced an ultimate magnification of 0.01$\times$ for images collected along the NF camera. 

\textbf{Intermediate Field (IF) Camera. } Behind the CRLs, we placed an alignment detector at a position 1.32-m behind the samples, with rotation along the $2\theta$ axis on its own separate rail system that allowed it to shift between the BF and DF beams. This intermediate-field detector used a 35-\um thick Ce:LuAG scintillator crystal with a Navitar zoom lens and an Allied Vision Manta g419b camera. Our intermediate-field camera had a resolution of 32-\um per pixel, corresponding to an optical magnification of 0.17$\times$. 

As is described fully in previous work \cite{Holstad2022,Poulsen2021}, the resolution function of DFXM enforces a strict Bragg condition, requiring very precise, accurate, and reproducible alignment. Due to the scarcity of XFEL facilities, no permanent and dedicated DFXM facility exists at this time; this necessitates full alignment of the entire microscope and the samples within the beamtime accessible for one experiment. We detail the steps required for efficient alignment that we perfected after experiments at PAL-XFEL and then at LCLS below: 
\begin{enumerate}[noitemsep,nolistsep]
    \item With the full 2D beam, align the direct beam to the FF detector to align, position, and calibrate the beam size \& optical magnification of the home-built BF detector.
    \item Align the CRL assemblies along the  $x_{\ell}$ and $y_{\ell}$ axes (dark-field, DF then bright-field, BF) using the IF alignment camera, with each lens stack placed along the direct-beam to avoid spatial jitter in the beam positioning during alignment.
    \item Watching on the FF camera, align the focus ($z$-position) of the CRLs for the BF arm using the TXM camera, then calibrate the TXM imaging system's resolution, field of view, and any aberration distortions with its full magnification.
    \item Align the 1D prefocusing CRLs to achieve the narrowest possible beam waist for the 1D vertical line beam at the sample position (calibrated using a TEM grid at the sample position). Calibrate beam size with magnified and unmagnified images (i.e. with and without the objective) along the direct beam. (Comparison of the magnified and unmagnified images allows one to calibrate the divergence of the XFEL beam)
    \item Orient the crystal of interest (or grain of interest in a polycrystal) into optimal Bragg condition along $\chi$,$\phi$,$\omega$ directions (intrinsic motor rotations of $R_y,R_x,R_z$ from bottom to top) using a photodiode, then a NF camera at the DF angle. 
    \item Orient the crystal's diffraction peak into the horizontal scattering geometry by rotating in along $\omega$, confirming its position on the NF, then IF, then FF detectors along the DF path.
    \item Rotate the laterally pre-aligned DF CRLs into the diffracted beam; adjust lateral alignment as needed for accurate positioning on the IF then FF cameras.
    \item Focus the DFXM imaging system by translating the DF CRL stack along the $z_i$ direction while monitoring on the DFXM FF camera. Once aligned, calibrate the resolution, field of view, and distortions on the DFXM images using a crystal with a TEM grid affixed to the exit surface of the crystal.
    \item For single crystals - translate the sample along $z_s$ to identify the upstream and downstream faces of the crystal, then select your ROI to ensure the motor encoders can track the ROI's precise position in the sample. 
    \item Rotate \& translate the sample as needed for data acquisition scans
\end{enumerate}

The specific points in this list have been identified with significant considerations, which we explain in more detail below:

\textbf{CRL alignment precision.} X-ray lenses are notoriously difficult to align due to their small effective apertures, long working distances, and the necessity of remote alignments. To align the 66-mm thick stack of imaging CRLs (280-\um pupil aperture) along the principal axis of the X-ray beam requires precise alignment along 5-axes - four lateral ($x,y,R_x,R_y$) and one along the beam's path for focusing ($z_i$) - and is highly sensitive \cite{Breckling2022}. Lateral alignment of DFXM CRLs is particularly challenging at XFELs because of the microscope's sensitivity; the high X-ray flux and strong interactions can thermally distort the material \cite{Chapman2010}. Since DFXM's high sensitivity to lattice distortions is amplified by the coherence and monochromaticity at XFELs \cite{Holstad2022}, even subtle thermal expansions and creep deformations can completely alter the material attributes resolved in DFXM images (by deflecting the diffraction condition). To simplify alignment, we thus performed lateral alignment for both the BF and DF CRL stacks along the more stable direct beam, using a custom circular carousel translation stage that independently rotates three separate carriages about a concentric $2\theta$ arc around the sample position. This allowed us to easily translate each lens stack between the diffracted and transmitted beams without changing the lateral alignment.

\textbf{Imaging axial strains} To position the FF detector along the diffracted beam is also rather difficult, as the sample-to-detector distance is 4-8 m propagation distance that must be precisely positioned along the diffracted beam, requiring specialized translation stages. Since the spot size of the beam is quite small, we identified that a temporary optical table for the DFXM far-field detector was very time consuming to position accurately. Our optimized design used the Large Angle Detector Mover (LADM) stage at XCS (LCLS) to position the FF detector along the 2$\theta$ arc of the diffracted beam. Long-term, detector stages like this will be essential for strain scanning capabilities in highly deformed crystals, which requires coupled translations of the CRL and FF-detector along the DFXM imaging arm to capture the full $2\theta$ content of the diffracting crystal.

\textbf{Alignment targets.} The specifics of the alignment targets must be carefully planned. Along the direct beam, we used standalone apertures and TEM grids to track the position, size, and distortions of the beam and imaging system. Optical magnifications of the detectors were measured with no sample, but with a TEM grid placed over the front face of the FF scintillator crystal to allow us to compare the grid to the known camera pixel sizes. We calibrated the magnification similarly with the full imaging system, placing the TEM grids instead at the sample position for the full calibration of the total magnification. For the DFXM, we calibrated the magnification and FOV using TEM grids placed on the exit surface of a diamond single crystal. While this only acts as a mask over the back face of the crystal, it does create a pattern on the imaged beam that is stretched along the same projection angle, enabling us to calibrate the distortions.  

\textbf{Precision of Scanning Stages.} For DFXM to accurately image the different points in reciprocal space, it must collect accurate and precise ($\sim$10$^{-5}$ radian,  $\sim$500-nm) scans of the sample along $x_{\ell}$, $\phi$, $\chi$, and $2\theta$, as shown in Fig.~\ref{fig:Scanning}. This requires scanning stages with high accuracy and precision \textit{along the required directions of travel for acquisition scans}. While many XFEL experiments take advantage of the continuous range of travel afforded by hexapods, the accuracy of travel through the entire scanning range is not as high as for classical stepper motors. We have thus found that accurate selection of the scanning motors is essential to the success of the ad- hoc (temporary) DFXM experiments at XFELs. For the best data acquisition, precise rotations along $\phi$, $\chi$ and $\omega$ require an extrinsic $\mu$ rotation stage at the bottom of the stack to align the $\omega$ rotation to be about the diffracting $\bf{Q}$ vector \cite{Poulsen2018}. Since it takes only two angular axes to rotate about the sphere, the designation of the $\chi$ and $\phi$ axes may be defined for each experiment (i.e. they are only a formalism).

\textbf{Beam stability. }The stability of the XFEL beam is important to this experiment because of the sensitive alignment of the microscope and the connection between goniometer angles and contrast mechanism. 
There are four aspects of the stability of XFEL beams: (1) the Poynting, (2) the spatial mode, (3) the timing, and (4) the spectrum. We detail these points below. 
\begin{enumerate}
    \item Deviations in the Poynting vector of the incident beam must be $<10\%$ of the beam's spatial profile; poynting stability is essential to define precise alignment of the $<500$-\um apertures of the CRLs \cite{Breckling2022}. For cases with poor spatial stability, we have used a large beam size with a smaller aperture, which converts the spatial instabilities to intensity instabilities.  
    \item Spatial mode stability is important to ensure that the interpretation of features in images arises from the material and not from the beam. This has been challenging for many XFEL imaging studies \cite{kudo2022x}, especially for experiments whose features of interest span a wide range of intensity gradients. The setup described in this work uses the simultaneous images from TXM to self-calibrate for spatial mode and intensity jitter known to be prevalent at XFELs. This effect may in some cases by mitigated by demagnifying the source, but can also be sensitive to aberrations along the undulator and X-ray optics halls. 
    \item While timing jitter is known to be a challenge at XFELs for sub-100-fs measurements \cite{vandriel2015timingtool}, the timing stability required for DFXM at XFELs usually would not require the use of those specialized timing tools. The time resolution required for each DFXM experiment is dictated by the physics of the material studied and specifically by the velocity of the dynamics as compared to the image's spatial resolution. For example, to image a 10 km/s wave would traverse a 0.4-\um pixel in 40-ps, dictating a timing stability of <40-ps would be required to image this wave with a spatial resolution of 400-nm.
    
    \item The spectral stability is perhaps the single most important point for DFXM image interpretability. Since $\lambda$ dictates the strain states that give rise to contrast in DFXM, the random fluctuations in the photon energy inherent to the source causes different $d$-spacings to diffract through the stationary lenses - changing the materials information inherent to the measurement. 
    The high brightness at XFELs stems from the Stimulated Amplification of Spontaneous Emission (SASE) process, which amplifies spontaneous emissions and therefore fluctuates its spectral content pulse to pulse across the entire range of the 0.3\% bandwith. This significant photon-energy jitter is greater than the $\gtrsim$18 eV required to entirely shift (in the present system) the $d$-spacing that accounts for the image out of the the 286-\um aperture of the imaging CRL, since $2\theta = f(\lambda,d)$. 
    
    In general, we find that DFXM experiments at XFELs require either a monochromator or a seeded monochromatic beam (i.e. filtering the SASE beam before amplification\cite{geloni2011selfseeding}) are best to ensure high sensitivity. 
    In highly deformed systems (e.g. shock waves, fracture), the very wide range of strain states present in the material would make it advantageous to use the full SASE spectrum to ensure that a sufficient range of strain states are represented to resolve the system. 
    Reversible systems (i.e. those that are feasible to measure with pump-probe experiments) may use signal averaging over many XFEL pulses to mitigate the spectral jitter, though this imposes an unchangeable minimum strain-resolution that would be set by the bandwidth of the SASE spectrum.
    
\end{enumerate}

\section{Results}

We present the results from our microscopy data in this section. To present the full overview of the data, we first summarize the raw data we measure on each detector, and explain what each image dataset tells us about the material and its dynamics. Next, we describe the analysis necessary to extract the full information contained in different types of sample rotational scans and explain how this changes our view of the material by imaging along different momentum vectors in reciprocal space, providing detail to explain the interpretation and analysis of the datasets. Finally, we present the results of rotational scans (rocking curve along $\phi$ and rolling scans along $\chi$) from our IF, and FF detectors, and explain the scientific insights afforded by each detector.

As discussed in the introduction, the timescales available with this instrument depend on the approach used to time- resolve the measurement. When probing reversible dynamics, the range of timescales accessible with XFEL-DFXM is from $\sim$ 50 fs through 1 ms, based on the tools for short-time calibration (minimum $\Delta$t) and the repetition rate of the XFEL (maximum $\Delta$t). To probe irreversible dynamics requires single-shot acquisitions in a bunch train, and is thus dependent on the inter-pulse spacing available at the XFEL and the detector readout rate. The range of lengthscales is more complex; the spatial resolution of the microscope is $\sim$1-\um in this work, but will likely be higher by using higher magnification focusing elements, as has been demonstrated at up to 30-nm resolution at synchrotrons \cite{kutsal2019esrf}. As noted in Beam Stability, the lengthscale of the information that must be resolved by the technique depends on the velocity and/ or rate of dynamics in the material being studied.

\subsection{Single-frame images with ultrafast high-resolution X-ray microscopy (U-HXM) instrument}
\begin{figure}[!ht]
    \centering
    \includegraphics[width=0.75\textwidth]{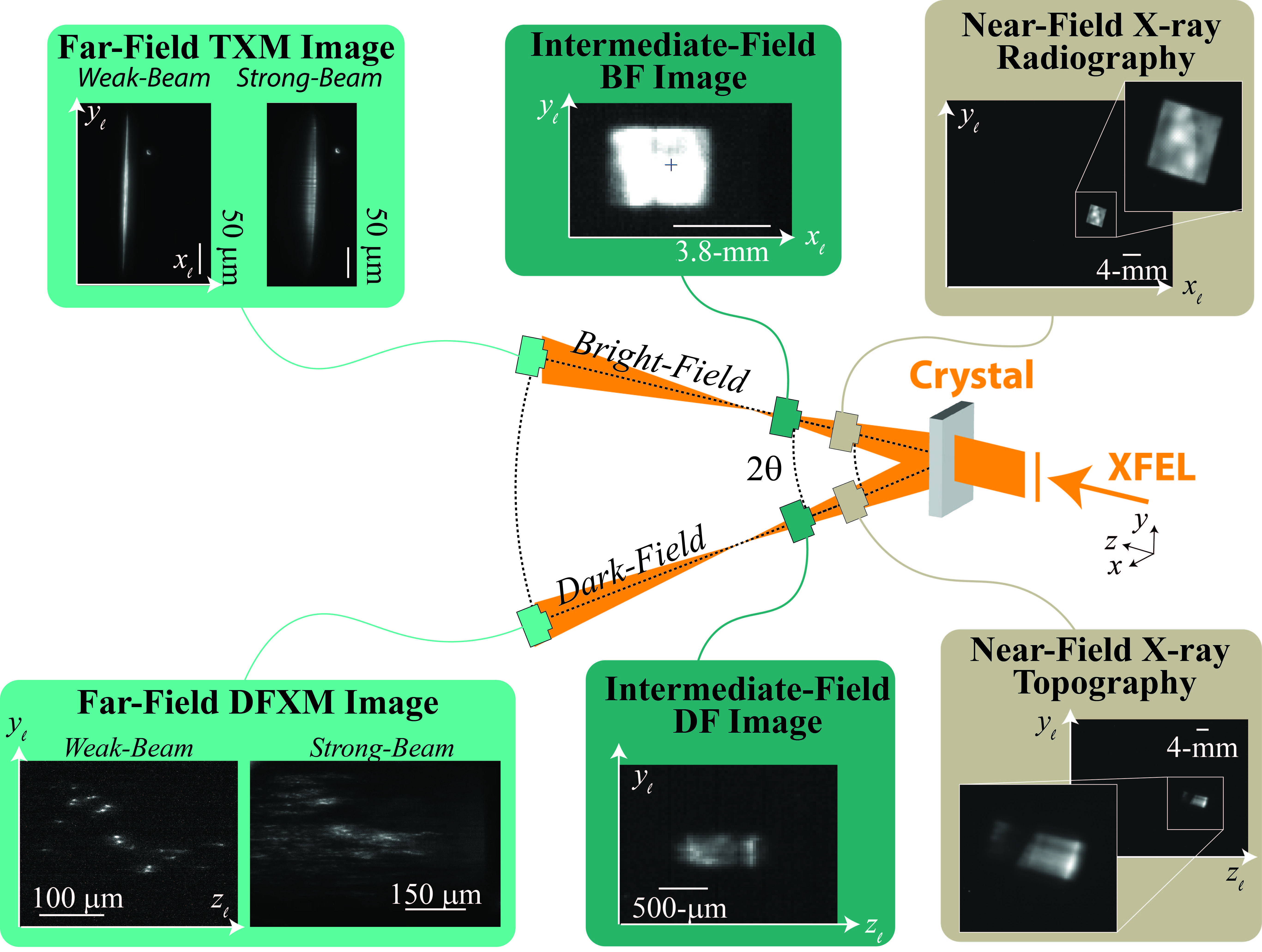}
    \caption{Sample images from each detector in the BF (above) and DF (below) imaging configurations. The general schematic for BF and DF are shown, with sample images collected in the NF (brown), IF (dark green) and FF (light green) shown for each position. NF and IF images are important for validation of the alignment of optics along the $(x_{\ell},y_{\ell})$ plane for BF to complement the $(z_{\ell},y_{\ell})$ observation plane along the DF imaging arm. The FF for each imaging line show representative images collected in the weak-beam and strong-beam conditions to illustrate the feature contrast (DF) and Bormann fan (BF), respectively. The BF images at the NF and IF positions show the 2D square beam, while the FF image for BF shows the line beam. We note that the IF image along BF is a screenshot and should thus only be considered qualitatively.}
    \label{fig:MicroscopeImages}
\end{figure}

For every pulse generated by the XFEL, the U-HXM instrument's data acquisition (DAQ) system saves the: (1) pulse energy, (2) goniometer stage positions, (3) TXM image of the line beam, and (4) DFXM image related to the elastic distortion states. As DFXM is a full-field microscope, each pulse produces a full real- space image on the camera; with beam tubes in place, we found that generally, a $\sim$1-mJ pulse energy gave DFXM and TXM images with appropriate signal-to-noise for single-shot acquisition. A full comparison of the different types of images collected along the BF and DF imaging arms is shown in Fig. \ref{fig:MicroscopeImages}, illustrating the change in resolution vs field of view afforded by each imaging modality.

The pulse energy measured from the upstream monitors informs the incident intensity required to normalize the signal intensity, while the goniometer angles catalogue the information required to interpret the microstructure and displacement states being sampled by each DFXM image. It is most effective to monitor the pulse energy with a monitor downstream of the monochromator due to the spectral jitter of the incident pulse (see Beam Stability section above). If the energy monitor is downstream of the monochromator, it may be used to calibrate the DFXM signal; if the intensity monitor is upstream of the monochromator, an in-line spectrometer is required, or conservation of intensity must be assumed across many averaged pulses between the BF and DF images in the FF. We note that future DFXM experiments may benefit from the new seeded monochromatic capabilities at XFELs, which produce low-bandwidth pulses with higher photon flux \cite{geloni2011selfseeding}.

As described above, images collected in the FF along the BF may comprise information on density variations but are also important to calibrate the DFXM images. As shown in Fig.~\ref{fig:MicroscopeImages}, the line beams collected from the BF camera can provide information about the spatial mode of the XFEL beam, the amount of dynamical diffraction occurring in DFXM, and context about the non-diffracting regions of the crystal (e.g. beam damage, phase transitions, etc.). In samples with no significant density variation, the TXM images can calibrate the fluctuations in the spatial mode of the XFEL, allowing one to correct for artifacts from the illumination. The TXM can also be helpful as a guide to quantify the amount of dynamical diffraction that occurs in a given diffraction condition (see TXM in Strong-Beam in Fig. \ref{fig:MicroscopeImages}). For nearly perfect crystals of sufficient thickness, dynamical diffraction (multiple scattering) causes the line-beam to scatter multiple times; for even numbers of scattering, the beam remains along the $\vec{k}_d$ wavevector, while an odd number of scattering events returns the beam back to the $\vec{k}_i$ wavevector. In both cases, the beam is translated from its initial position due to these multiple scattering events. This effect, termed the Borrmann effect \cite{saccocio1965borrmann}, makes the TXM line beam spread out laterally, as ghost-fringes appear next to the primary beam. Information about the extent of dynamical diffraction can be helpful in understanding how to interpret image features with DFXM, as they also complicate signal in DFXM images. Finally, the TXM images give important information about the structure and its evolution in the entire illuminated region of the sample. Notably, DFXM is designed to  image the specific part of the crystal that gives rise to scattering over a narrow range of reciprocal space, giving it very high resolution in real and reciprocal space. In general it does not provide any information on other material phases and grains that are oriented differently, as there is either no diffraction signal or the diffraction occurs at angle not compatible with the DFXM settings. With simultaneous density contrast that images these ``missing regions,'' TXM gives a complementary view of the material, giving important context for interpretation of the DFXM images. At the PAL-XFEL, we also measured simultaneous wide-angle X-ray diffraction on the side opposite from
the DF beam (mirrored across the BF beam), however, we do not include this in the presentation in this work, as that data was not
meaningful without a phase transition or breakup of the crystal grains to observe. In future experiments, simultaneous diffraction
can be helpful for interpretation of multi-phase and/or polycrystalline changes to the material.

As described in Poulsen, et al. \cite{Poulsen2021}, each raw DFXM image contains bright and dark features that map specific components of the crystallography, detailing components of the strain and orientation states along the lattice plane being measured, $hkl_i$. The content of these images is related to the spatial derivatives of the elastic displacement fields, which includes information about defects in materials and distortion waves like phonons, shock waves, and heat \cite{Holstad2022b}. 

A full comparison of the different types of images collected along the bright-field (BF) and dark-field (DF) imaging arms is shown in Figure \ref{fig:MicroscopeImages}, illustrating the change in resolution vs field of view afforded by each imaging modality. 

\subsection{Overview of DFXM scanning modalities}
While significant insights may be gathered by time-resolving single-frames with DFXM, full information about dynamics of interest with DFXM, require scans along the parameters outlined in Table \ref{tab:DFXM_scans}. A full DFXM scan is collected by measuring images at a series of different motor positions, photon energies, and/or times as described in Table \ref{tab:DFXM_scans}. Strain and orientation scans probe different populations of reciprocal space in a region around the diffraction vector; alternatively this is presented in terms of axial strain and orientation (a local pole figure). The 3D and time scans capture the same structural information as  a function of position and time, respectively (without changing the contrast mechanism). In most experiments, a combination of these scans are performed, where some combination of structure and dynamics are measured at a series of positions across different axes of a crystal's diffraction extent (rocking or rolling curve).

\begin{table}[h]
    \centering
    \begin{tabular}{ | >{\raggedright\arraybackslash}m{3cm} | >{\centering\arraybackslash}m{2cm}| >{\centering\arraybackslash}m{2cm} | >{\centering\arraybackslash}m{2cm} | >{\centering\arraybackslash}m{2cm} | } 
    \hline
         & \bf{Axial Strain} & \bf{Orientation} & \bf{3D Structure} & \bf{Time} \\ \hline
         \bf{Scan Parameters} & $2\theta$ or $\lambda$ & $\phi$ and/ or $\chi$ & $\omega$, and $x_{\ell}$ & $t$ or $\tau_{pp}$ \\ \hline
    \end{tabular}
    \caption{Scanning modalities of DFXM that give observable image contrast for different types of structural information about material dynamics.}
    \label{tab:DFXM_scans}
\end{table}

For strain and orientation scans, each image spatially maps the populations of displacement fields that meet the Bragg condition at that position in reciprocal space, within the reciprocal- space resolution function of the microscope (as shown in Fig.~\ref{fig:Scanning}b). As described fully in Jakobsen, et al,\cite{Jakobsen2019} images collected at the brightest point of a rotational scan display the entire extent of the undeformed grain of the crystal domain, and define the \textit{strong-beam condition}. Images collected at microscope settings at the edges of the rotational scan only describe specific anomalous types of distortions in the lattice, showing the defects or other sparse structural features in the sample; these images define the \textit{weak-beam condition}. The nomenclature in this terminology was selected  in analogy to dark-field TEM experiments in electron microscopy. At XFELs, the time-resolved scanning capabilities introduce another set of displacement states not described in previous DFXM work: \textit{temporally weak-beam states}. We use this term to denote new weak-beam distortion states that populate during the transient dynamics probed by the experiment, but that are not populated at $t < 0$, before the pump pulse.

Strain scans are typically acquired at the synchrotron by collectively moving the detector and CRL stages coherently along the $2\theta$ arc of the diffracting crystal. While the dedicated instruments at synchrotrons make this approach viable, the extra setup requirements at XFELs make it simpler to hold the entire experiment stationary, scanning the $d$-spacing with small changes to the photon energy for $\Delta\lambda$. For the case of probing diamond (111) planes by an X-ray pulse of $E$ = 10.1 keV and $\Delta{E}/{E} = 10^{-4}$, a single X-ray pulse with a bandwidth of $\Delta{E} = 1$ eV covers a narrow range of  $\delta{d} = 2\times10^{-5}$-nm and $2\delta\theta = 0.004^{\circ}$. For example, in the present experiment, a strain scan spanning a range of $\Delta\varepsilon = 1\times10^{-3}$ ($\Delta{d} = 2\times10^{-4}$ nm) would require 10 steps of $\Delta E= 1$ eV. This range of strain scanned by changing photon energy is equivalent to that scanned by changing scattering angle for $2\Delta\theta = 0.03^{\circ}$.  Strain scans were not performed in this work, but we include this description for future
users' benefit in planning experimental capabilities.

The reciprocal space resolution function for DFXM is highly anisotropic, with the beam divergence accounting for the highest sensitivity in $\vec{q}$; this effect is amplified by the coherence of the XFEL \cite{Holstad2022}. To map different strain components, it is important to consider the orientation of the lattice in relation to the axis of the resolution function, as is discussed fully in \cite{Poulsen2018,Poulsen2021}, to ensure the appropriate match between the microscope's resolution and material's displacement fields. 

Different types of dynamics are more suitable to study in the strong-beam or weak-beam condition based on the grain sizes and the spatial population of states present. In past, we have found defect measurements in low-defect density materials are easiest to interpret in the weak-beam conditions. For highly deformed crystals, the local orientation changes are typically much larger than the Darwin width and the diffracted signal can be interpretable when acquired in the strong-beam condition. At the XFELs, the temporally-weak condition will be most helpful to study dynamics in many samples and conditions. While the temporally-weak beam condition is usually the easiest data to interpret during an experiment, however, those data often lack sufficient information about the initial microstructural features in the sample that
inform how and why those dynamics began. As such, we suggest a balance between strong, weak, and temporally-weak conditions when planning for XFEL-DFXM studies. 

\begin{figure}[h]
    \centering
    \includegraphics[width = 0.9\textwidth]{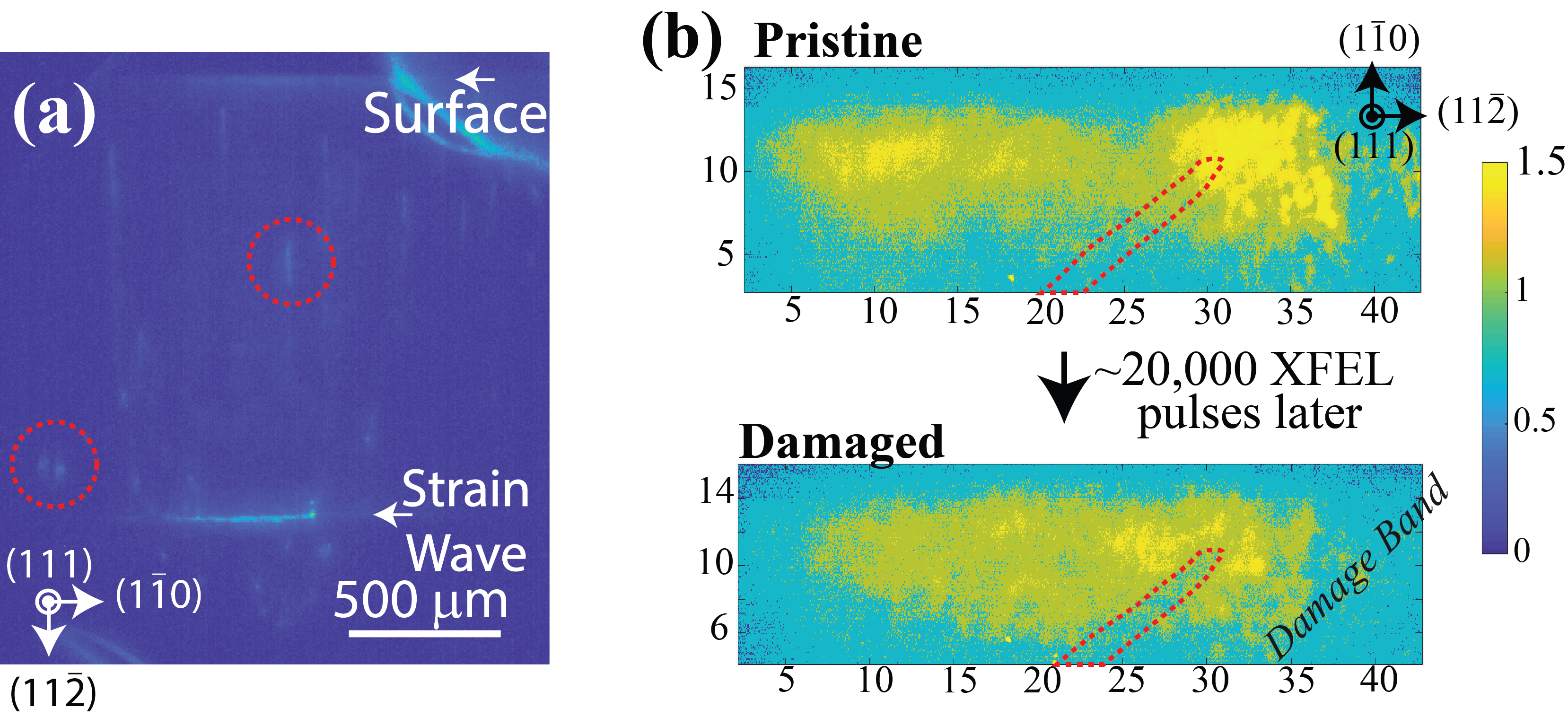}
    \caption{Raw data showing DFXM images captured at (a) LCLS using pump-probe acquisition mode, and (b) the PAL- XFEL using real-time image acquisition modalities. The image in (a) shows an integrated image of a strain wave traversing a diamond single-crystal, while (b) shows the incipient beam damage after 20,000 pulses from the XFEL at PAL-XFEL. We note that the crystallographic legends shown in all images are rotated by an angle of $\theta=17.5^{\circ}$ about the vertical axis from the orientations shown. For description of the dynamics of the strain wave images and measurements, see Holstad et al. \cite{Holstad2022b}.}
    \label{fig:SingleFrames}
\end{figure}

When studying dynamics at the XFEL, it is important to consider the tradeoffs between different approaches. For irreversible processes, real-time images must be collected to capture the dynamics of a material as it changes - this approach will always be limited by the camera acquisition speeds (frame rate and shutter speed), the inter-pulse separations for timing sequences, and the data transfer rates that are possible for the data acquisition system. Optical fibers for rapid data transfer are essential to achieve sufficient data transfer rates for the high-resolution and high dynamic range detectors required to acquire the full content of DFXM images ($>$12 bit-depth per pixel). From our experience, the U-HXM instrument performs best when each camera is connected to its own data-transfer line to enable parallel data transfer capabilities. From our experience, the $10^{12}$ photons per pulse available at the XFEL are indeed able to acquire sufficient signal-to-noise for single-shot acquisitions, however, for weak-beam conditions, this can introduce significant challenges in interpretation (e.g. Fig.~\ref{fig:SingleFrames}b). Our experiments at the PAL-XFEL observed damage bands appearing in diamond after 20,000 pulses of the XFEL ($\sim$14.3 Gy of radiation absorbed, assuming 0.1-mJ/pulse), however, the low intensity at the DFXM detector presented challenges in distinguishing this signal from burns to the Kapton tape behind it. 

By contrast, for crystals exhibiting reversible dynamics, a pump-probe modality may be employed. In that modality, an optical pump laser drives reversible dynamics in the same sample region illuminated by the XFEL probing laser, at a series of time delays, $\tau_i$, that can be stepped through the full dynamics of the system. This approach enables signal averaging over hundreds or thousands of pulses, giving results like the image in Fig.~\ref{fig:SingleFrames}a, offering much higher signal to noise. The pump- probe approach often requires longer than anticipated acquisition times, as the XFEL pulse may require attenuation to avoid sample damage over the long duration scans. The example image in Fig.~\ref{fig:SingleFrames}a showing a strain wave propagating in diamond, generated by laser irradiation of the gold surface transducer layer as discussed in \cite{Holstad2022b}.

Finally, as presented above, one DFXM image probes a single observation plane in the specimen. A spatial 3D map may therefore be acquired by collecting image stacks while stepping the sample along the $x_{\ell}$ axis, though this was beyond the scope of our initial
study. We note that the sensitivity of the DFXM microscope requires that the images in such $(x,y,z)$ scans are collected with rotational scans of at least one tilt axis for each observation plane to correct for motor drift ($>0.0005^{\circ}$) or spatial changes to the microstructure over the scan volume \cite{yildirim2022}. As the reciprocal space resolution function is thinnest in direction $\phi$, it is natural to perform this additional scan along $\phi$. 

As an alternative to layer-by-layer mapping, the condenser may be removed an the specimen therefore illuminated by a square shaped beam. The DFXM images then represent a projection, though along a more complex geometry due to the tilted angle of the
projection.  Similar to classical tomography, 3D mapping may then be facilitated in a tomographic fashion by rotating in $\omega$ around the diffraction vector. This so-called topo-tomographic DFXM modality can give faster 3D information but is more complicated in terms of sampling \cite{Simons2015}.  

In large crystalline materials, combined analysis of the NF, the IF and FF images is quite helpful. The large difference in the field of view provides context as the NF and IF detectors can capture  global dynamics occurring in volumes that are outside the field-of-view of  DFXM (over mm- lengthscales). Changes to the real-space sample (e.g. thermal expansion, fatigue, etc.) would be evident in the NF and IF cameras, showing images that clearly demonstrate a change in the strong-beam image. New components of reciprocal space that populate are most evident at the IF camera, where satellite images may begin to appear to indicate new strain/orientation components that could also be of-interest to DFXM but would require realignment to the new $2\theta$ and $\eta$ positions. In our experiment, we observed images along the IF that
showed the strong-beam and the temporally-weak beam signals to identify the position of the temporally weak-beam signal (image
not saved).

\subsection{Analysis for U-HXM Data}
In this section, we provide an overview of the data analysis approach to evaluate the crystalline domains sampled by different scan modalities. For a detailed overview of the analysis, we encourage the reader to explore more thorough guides of specific analysis tools developed in darfix \cite{Ferrer2022darfix} and elsewhere \cite{yildirim2022}.

As in classical XRD, the diffraction condition in DFXM is determined by the conservation of momentum criteria that arise from the sum of the momentum vector for the incident, $\vec{k}_{i}$, and diffracted, $\vec{k}_d$, beams being equal to a lattice vector, $\vec{Q}$. While a perfect crystal would diffract only at a specific value of $\vec{Q}_0$, the defects and distortions within the lattice cause additional $\vec{Q}_i$ states to populate as well, mapping the vectors characteristic of the subtle distortions in the lattice spacing and orientation, i.e., the displacement gradient tensor field \cite{Poulsen2004}. The specific components of the displacement gradient tensor field that contribute to the diffracted signal arise from the crystal's orientation and the position of the detector. 
For a full derivation of the contrast mechanism for DFXM, we refer the reader to recent work using ray optics \cite{Poulsen2021} and wavefront propagation \cite{Carlsen2022} with synchrotron radiation, and XFEL sources \cite{Holstad2022}.  

Experimentally, DFXM crystallographic scans are collected by capturing images produced by the crystal when it is rotated to each point along the rocking curve. The corresponding stack of images then has coordinates of $(z,y,\chi,\phi, 2\theta)$ (assuming $\omega=\mu=0$), and may be reconstructed into a single mosaicity map by constructing an image for which each pixel ($z_i,y_i$) has an intensity, $I(z_i,y_i)$, that corresponds to the maximum intensity observed at that pixel throughout the scan, $I(z_i,y_i) = \text{max}_{\chi,\phi,2\theta}(I(z_i,y_i,\chi,\phi))$, and is plotted with a color, $C(z_i,y_i)$, that encodes the $\chi,\phi,2\theta$ orientation that diffracted with that maximal intensity. A schematic is shown in Fig.~\ref{fig:Scanning} displays the scanning procedure and associated image sequences, with experimental data to show the characteristic images and reconstructed results (collected at the ESRF).

We note that the above description of the Center-of-Mass (COM) images is relevant when measuring the average properties of the lattice locally, e.g. measuring the orientation of a local domain. When studying the distortion fields surrounding single defects, the average properties typically wash out the lower intensity signal from defect structures, causing single images of the weak-beam conditions to be preferable \cite{Jakobsen2019,yildirim2022}.

For static or reversible systems, the detailed mapping accessible from reconstructed DFXM scans along $\phi$, $\chi$, and $2\theta$ would be most important to fully characterize unknown defect structure that is not well understood. For irreversible systems (e.g. fracture, radiation damage), a crystal grain with sufficiently well-understood defect structures may be measured in real-time, using simulations of all possible defect structures to evaluate the image features appearing in DFXM images collected at only a single crystal orientation. This may be done manually \cite{DresselhausMarais2021}, or using Bayesian inference for physics-informed image interpretation \cite{brennan2022analytical}.

\subsection{Guide for Image Reconstruction from DFXM Scans}
 In this section, we introduce the general approach we used for the image processing treatments, and reconstructions to measure the observables for the basic rotational scans. 

\textbf{Noise Removal. }After passing through the monochromator, the photon-energy fluctuations from the SASE beam manifest as 100\% fluctuations in the XFEL beam intensity. Thus, for pump-probe and static experiments, we average over multiple X-ray pulses at a fixed sample position to improve the signal-to-noise ratio for the image at the selected microscope conditions. During the experiment, dark frames with the X-ray and optical laser beams blocked were acquired and saved every 13 frames to measure the pedestal noise from the detector \cite{Legge1987pedestal}. 
For the XFEL beam, we account for the intensity variation on a shot-to-shot basis by integrating DFXM and TXM images over all images for a given microscope condition.  For sufficiently sampled scans, this
may be calibrated by comparison of the integrated intensity from each microscope position in both the DFXM and BFXM
images, plotted across the rocking curve. By then comparing the integrated intensity vs microscope condition for the full scan and fitting them to a Gaussian (DF) and an inverted Gaussian (BF),
we plotted the two complementary rocking curves measured for the DFXM (i.e. $I_{DF}(\phi)/I_0 \approx
1-I_{BF}(\phi)/I_0$). If the BFXM shows a complementary Gaussian
curve to the DFXM rocking curve, this confirms that intensity loss arose from only the change in intensity of the XFEL beam,
and the multiplier may be deduced from curve-fitting of the integrated intensity rocking-curve functions. Because of the noise subtraction, there were pixels with a negative intensity, requiring that all negative values be thresholded to 0 before subsequent processing. 
We note that when plotting these images, it is important to ensure that the minimum and maximum color limits are set carefully, as hot pixels in the image make python/matlab choose poor colormap limits if not set manually. The 1$^{\text{st}}$ and 99$^{\text{th}}$ percentile are normally a good choice for the limits of the colormap.

\textbf{Scanning Reconstructions}. A powerful way to reconstruct the strain or mosaicity fields from the scans collected by DFXM is to  compile a COM data reduction over one or more scanning axes considered. Before starting the comparison between different motor positions of the scan, we first normalize the summed images to account for the intensity variation of the XFEL. Noise in the images collected at angles far from the center of the rocking curve have an outsized impact on the center of mass (COM) calculation, and will bring the COM towards the center of the scan. We reduce the impact of this by setting the intensity of pixels below the $n^{\text{th}}$ percentile to zero, or alternately subtracting a fixed value from all pixels before normalization.
Finally, we then calculate the center of mass by fitting the $I_{x,y}(\phi,\chi,2\Delta\theta)$ function for each pixel to a Gaussian and taking the weighted average of each pixel for each scanning parameter, as mentioned in the Analysis for U-HXM
data section.. More detailed descriptions of how the strain and mosaicity are solved from this approach can be found elsewhere.\cite{Simons2015}

\subsection{Example Image Reconstructions}
\begin{figure}[!ht]
    \centering
    \includegraphics[width=0.65\textwidth]{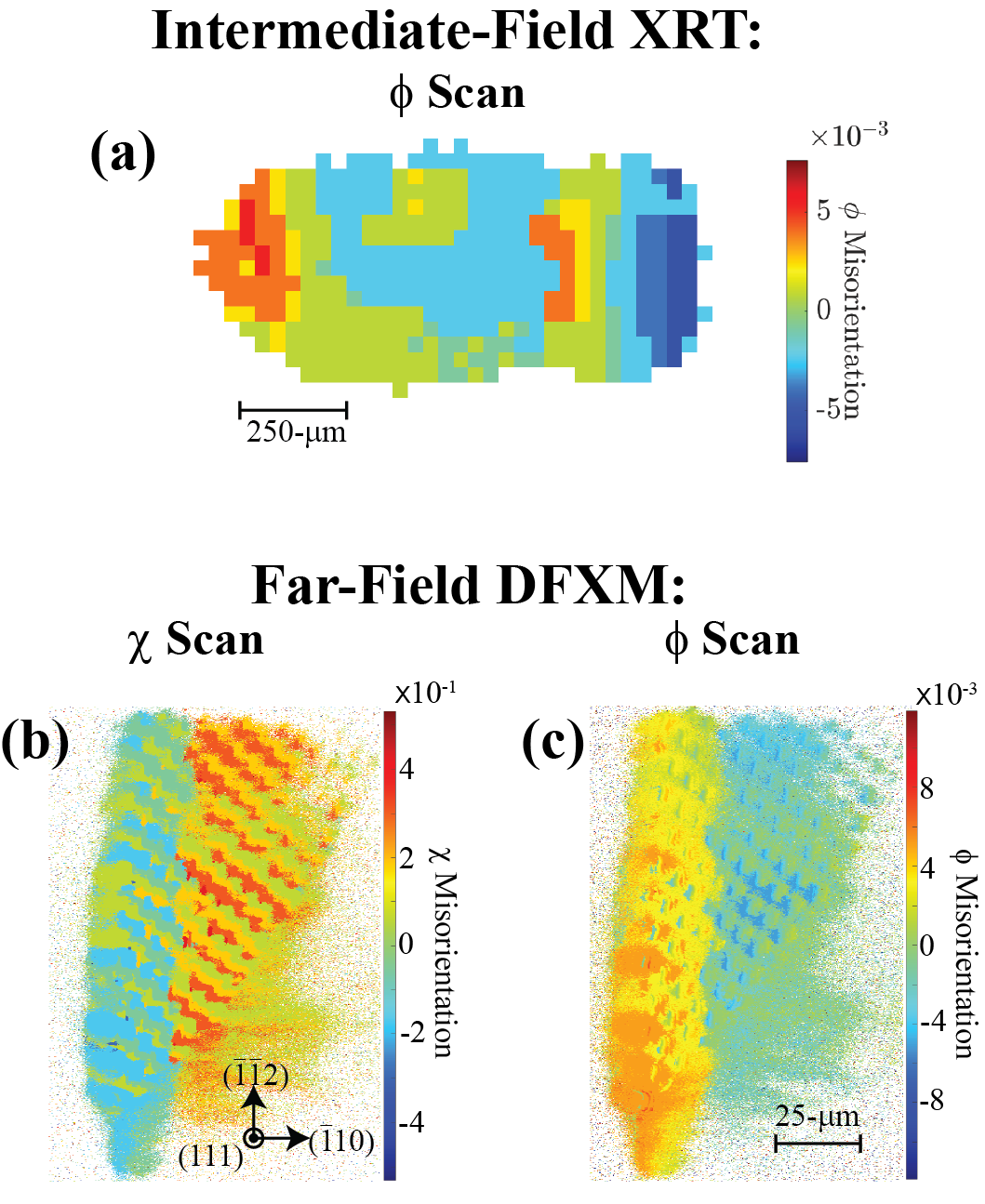}
    \caption{Reconstructed diffraction-contrast imaging scans, collected (a) at the IF detector, producing a COM XRT image reconstructed from a $\phi$, and (b-c) after magnification at the FF detector, COM reconstructions along (b) $\chi$ and (c) $\phi$. Scale bars for the DFXM images are identical, while the XRT image is 10$\times$ larger, but the crystallographic legend are consistent for all images.}
    \label{fig:DFXM_Scans}
\end{figure}

To demonstrate the mapping of local orientation, we include set of representative reconstructed images in Figure~\ref{fig:DFXM_Scans}, showing the relationships between the XRT and DFXM images as well as the different types of scanning modalities. Fig.~\ref{fig:DFXM_Scans}(a) shows the reconstruction from an unmagnified XRT image series, while (b-c) show a stretched TEM grid over a domain wall in diamond \{111\}, measured along the $R_y$ and the ($R_x,R_z$) rotation stages to map $\phi$ and $\chi$, respectively.

The XRT image shown in Fig.~\ref{fig:DFXM_Scans}a illustrates the additional context provided by the IF scans. Fig.~\ref{fig:DFXM_Scans}a shows the reconstructed image of a $\phi$ scan collected from an $R_y$ scanning acquisition. The $\sim1150$-\um field of view along the horizontal axis shows the entire extent of the observation plane in the crystal that is viewed in our images, $\left( \frac{D}{\sin(2\theta)}\text{ for }D=660\text{\um crystal thickness}\right)$, as viewed by diffraction- contrast microscopy. The XRT image is highly pixelated, showing a significantly lower spatial resolution, but maps the contextual structure of the crystalline domain across the entire extent of the domain (that passed through the CRL). In this way, XRT provides context at the IF detector for the full crystal information - guiding the imaging alignment and the interpretation of the microstructure in the DFXM images that follow. 

Fig.~\ref{fig:DFXM_Scans}(b-c) show the representative images that may be collected in the FF in the magnified DFXM geometry, showing rotational scan reconstructions from the $\chi$ ($R_x,R_z$) and the $\phi$ ($R_y$) directions, respectively. Both images map the exact same region of the diamond (though a different one than in XRT) - with a low-angle boundary bisecting the left and right regions of the crystal by an angle that maps between the two rotational directions. While Fig.~\ref{fig:DFXM_Scans}(c) shows the $R_y$ rotations that map directly to the $\phi$ rotational axis described above, the reconstructed results in Fig.~\ref{fig:DFXM_Scans}(b) included rotational offsets in $R_z$ to correct for the fact that the $R_x$ rotation shown in Fig.~\ref{fig:Scanning} includes components of rotation in $\chi$ and $\omega$. The $R_z$ corrections were performed by ensuring that the same TEM-grid fiducial appeared over the same pixels in each frame of the rotational scan. As such, the image in Fig.~\ref{fig:DFXM_Scans}(b) approximates a $\chi$ scan. 

The results from Fig.~\ref{fig:DFXM_Scans}b-c indicate that the misorientation of the diamond across the boundary in its center is $\sim5\times10^{-1}$ degrees in the $\chi$ direction, and $-6\times10^{-3}$ degrees in the $\phi$ direction. More thorough investigation of this type of feature may inform the Burgers vector and spacing of the dislocations that comprise the boundary - serving as a direct measurement of the microstructure.

\section{Discussion}

Our results in this work demonstrate a new capability for X-ray science that holds opportunities to enhance DFXM from the 0.1 second timescale all the way to the 100- fs timescales. Using the instrument design we describe here, this setup allows for simultaneous multi-beam imaging experiments that can probe multiscale dynamics in materials under a range of different conditions. Using the simultaneous DF and BF imaging (DFXM and TXM, respectively), we have demonstrated a new opportunity to explore the reciprocal space components that underlie a specific feature captured by a TXM image of the material. This is important in applications like fracture or amorphization, in which materials begin to ``cloud'' in an image collected along the direct-beam because the transmitted beam attenuates from Fresnel and/or Bragg scattering as voids begin forming that instigate the fracture process \cite{Li2011}. 
While the BF TXM spatially maps a material’s density by imaging along the transmitted beam (magnified version of radiography), DFXM resolves the local lattice distortions that underlie those density variations. Together, these techniques hold the opportunity to resolve snapshots of how defects and domains (and their associated strain fields) grow, propagate and interact as they cascade into large-scale material dynamics. This work sets the stage for a new host of experiments to study how mesoscale defects deep beneath a material’s surface initiate phase transitions, strengthen materials, modify their physical properties, and beyond, as they respond to external stimuli. 

Beyond the impact of this technique, we also stress the sensitivity of this microscope, and the importance to consider the full scope of preparations required to effectively sample materials with DFXM. We note that effects like dynamical diffraction and phase transformations can introduce uncertainties to the  interpretation of the data in ways that require complementary probing methods to be able to interpret. We anticipate that future theory papers may explore this interpretation explicitly, enabled by the additional coherence available at XFELs. We additionally comment that our choice of CRLs as the focusing optics for the microscopes also
holds opportunity for future directions. To capture enough of reciprocal space to avoid imaging artifacts, DFXM requires the largest
possible numerical aperture, which is limited by the pupil aperture available with X-ray lenses. We selected CRLs as the best
compromise between flexible effective focal lengths, relatively high pupil apertures, relatively high sample-to-lens distances, and
ease of alignment. We note that alternative focusing optics have been explored in the synchrotron DFXM and are a viable future
option for XFEL DFXM.

Finally, we also emphasize that beyond the challenges to building a DFXM instrument from scratch, there are additional considerations required for these experiments at XFELs due to the unique attributes of the source. In the sections above, we have detailed how the XFEL's polarization, photon energy jitter, and nonuniform spatial modes can add complexity to interpretation if not properly controlled during the experiment. In addition, we also emphasize the importance of considering the damage threshold of the material carefully when designing experiments. Our work did not need to consider beam damage for the pump-probe and static measurements because we used diamond samples for the proof-of-concept measurements, which has the highest known X-ray damage threshold. For other samples, the material's damage threshold must be compared against the power \textit{density} of the X-ray beam. As DFXM often is used to resolve dynamics in long-range microstructural features, we often enlarge the microscope's field of view to lower the power density, however, attenuators may be used for sufficiently high efficiency detectors and/or highly scattering samples.

The use of DFXM at XFELs offers new insights into mesoscale materials science, offering complementary information to the current state of the art in XFEL imaging. Current XFEL imaging along the X-ray diffracted beam has been performed for X-ray ptychography and Bragg XCDI \cite{martin2012femtosecond}. While X-ray ptychography is often used at synchrotrons to enhance the resolution of XCDI, at XFELs it has been employed primarily for full nano and micro beam characterization\cite{schropp2013full,daurer2021ptychographic}. Coherent diffraction imaging has been used extensively at XFELs to image ultrafast structural dynamics in nanomaterials, including acoustic phonons in zinc oxide\cite{ulvestad2017bragg}, melting in nanoparticles\cite{Clark2015}, and deformations in catalysts\cite{kang2020bcdi_catalysis}. At synchrotrons, both techniques are leveraged for providing both the amplitude and phase of the diffracted beams, which can give more subtle information about phase-contrast from fine features. Ptychographic analysis and reconstruction was recently extended to DFXM studies\cite{carlsen2022fourier}, and may be able to provide this type of information in future implementations of XFEL-DFXM.

\section{Conclusion}

We have presented a full integrated design for multi-modal ultrafast imaging at XFELs using simultaneous DFXM and TXM. Our new technique provides opportunities to resolve dynamics that occur from slow to ultrafast timescales – spanning 13 orders of magnitude beyond the current state of the art for DFXM dynamics. The shift from stroboscopic integrated measurements to the single-shot ultrafast measurements that are now accessible at XFELs can enable a new class of experiments in materials science. We envision this new capability offers opportunities to understand ultrafast deformations (e.g. fracture), rapid phase transformation heterogeneity, thermal transport, and charge-density waves in superconductors, and important phenomena in many other fields.

\bibliography{references}

\begin{thebibliography}{10}
\urlstyle{rm}
\expandafter\ifx\csname url\endcsname\relax
  \def\url#1{\texttt{#1}}\fi
\expandafter\ifx\csname urlprefix\endcsname\relax\def\urlprefix{URL }\fi
\expandafter\ifx\csname doiprefix\endcsname\relax\def\doiprefix{DOI: }\fi
\providecommand{\bibinfo}[2]{#2}
\providecommand{\eprint}[2][]{\url{#2}}

\bibitem{Bulatov2006}
\bibinfo{author}{Bulatov, V.~V.} \emph{et~al.}
\newblock \bibinfo{journal}{\bibinfo{title}{Dislocation multi-junctions and
  strain hardening}}.
\newblock {\emph{\JournalTitle{Nature}}} \textbf{\bibinfo{volume}{440}},
  \bibinfo{pages}{1174–1178} (\bibinfo{year}{2006}).

\bibitem{Zhao2010}
\bibinfo{author}{Zhao, K.}, \bibinfo{author}{Pharr, M.},
  \bibinfo{author}{Vlassak, J.~J.} \& \bibinfo{author}{Suo, Z.}
\newblock \bibinfo{journal}{\bibinfo{title}{Fracture of electrodes in
  lithium-ion batteries caused by fast charging}}.
\newblock {\emph{\JournalTitle{Journal of Applied Physics}}}
  \textbf{\bibinfo{volume}{108}}, \bibinfo{pages}{073517}
  (\bibinfo{year}{2010}).

\bibitem{tuller2011pointdefects}
\bibinfo{author}{Tuller, H.~L.} \& \bibinfo{author}{Bishop, S.~R.}
\newblock \bibinfo{journal}{\bibinfo{title}{Point defects in oxides: tailoring
  materials through defect engineering}}.
\newblock {\emph{\JournalTitle{Annual Review of Materials Research}}}
  \textbf{\bibinfo{volume}{41}}, \bibinfo{pages}{369--398}
  (\bibinfo{year}{2011}).

\bibitem{Zepeda-Ruiz2017}
\bibinfo{author}{Zepeda-Ruiz, L.~A.}, \bibinfo{author}{Stukowski, A.},
  \bibinfo{author}{Oppelstrup, T.} \& \bibinfo{author}{Bulatov, V.~V.}
\newblock \bibinfo{journal}{\bibinfo{title}{Probing the limits of metal
  plasticity with molecular dynamics simulations}}.
\newblock {\emph{\JournalTitle{Nature}}} \textbf{\bibinfo{volume}{550}},
  \bibinfo{pages}{492–495} (\bibinfo{year}{2017}).

\bibitem{armstrong2021dislocation}
\bibinfo{author}{Armstrong, M.~D.}, \bibinfo{author}{Lan, K.-W.},
  \bibinfo{author}{Guo, Y.} \& \bibinfo{author}{Perry, N.~H.}
\newblock \bibinfo{journal}{\bibinfo{title}{Dislocation-mediated conductivity
  in oxides: Progress, challenges, and opportunities}}.
\newblock {\emph{\JournalTitle{ACS nano}}} \textbf{\bibinfo{volume}{15}},
  \bibinfo{pages}{9211--9221} (\bibinfo{year}{2021}).

\bibitem{Goldsmid2001}
\bibinfo{author}{Goldsmid, H.} \& \bibinfo{author}{Nolas, G.}
\newblock \bibinfo{title}{A review of the new thermoelectric materials}.
\newblock In \emph{\bibinfo{booktitle}{Proceedings ICT2001. 20 International
  Conference on Thermoelectrics (Cat. No. 01TH8589)}}, \bibinfo{pages}{7246523}
  (\bibinfo{organization}{IEEE}, \bibinfo{year}{2001}).

\bibitem{Russell2005}
\bibinfo{author}{Russell, A.~M.} \& \bibinfo{author}{Lee, K.~L.}
\newblock \bibinfo{title}{Defects and their effects on materials properties}.
\newblock In \bibinfo{editor}{Russell, A.~M.} \& \bibinfo{editor}{Lee, K.~L.}
  (eds.) \emph{\bibinfo{booktitle}{Structure-Property Relations in Nonferrous
  Metals}}, chap.~\bibinfo{chapter}{2}, \bibinfo{pages}{18--27}
  (\bibinfo{publisher}{John Wiley \& Sons, Inc.}, \bibinfo{address}{Hoboken,
  NJ}, \bibinfo{year}{2005}).

\bibitem{kubin1996dislocation}
\bibinfo{author}{Kubin, L.~P.}
\newblock \bibinfo{title}{Dislocation patterns: experiment, theory and
  simulation}.
\newblock In \emph{\bibinfo{booktitle}{Stability of materials}},
  \bibinfo{pages}{99--135} (\bibinfo{publisher}{Springer},
  \bibinfo{year}{1996}).

\bibitem{ross2019tem}
\bibinfo{author}{Ross, F.~M.} \& \bibinfo{author}{Minor, A.~M.}
\newblock \bibinfo{title}{In situ transmission electron microscopy}.
\newblock In \emph{\bibinfo{booktitle}{Springer Handbook of Microscopy}},
  \bibinfo{pages}{101--187} (\bibinfo{publisher}{Springer},
  \bibinfo{year}{2019}).

\bibitem{smith2013studies}
\bibinfo{author}{Smith, G.}, \bibinfo{author}{Hudson, D.},
  \bibinfo{author}{Styman, P.} \& \bibinfo{author}{Williams, C.}
\newblock \bibinfo{journal}{\bibinfo{title}{Studies of dislocations by field
  ion microscopy and atom probe tomography}}.
\newblock {\emph{\JournalTitle{Philosophical Magazine}}}
  \textbf{\bibinfo{volume}{93}}, \bibinfo{pages}{3726--3740}
  (\bibinfo{year}{2013}).

\bibitem{yildirim2022}
\bibinfo{author}{Yildrim, C.} \emph{et~al.}
\newblock \bibinfo{journal}{\bibinfo{title}{Extensive 3d mapping of dislocation
  structures in bulk aluminum}}.
\newblock {\emph{\JournalTitle{arXiv:2208.14284}}} \bibinfo{pages}{1--14}
  (\bibinfo{year}{2022}).

\bibitem{Bacon}
\bibinfo{author}{Hull, D.} \& \bibinfo{author}{Bacon, D.~J.}
\newblock \emph{\bibinfo{title}{Introduction to Dislocations}}
  (\bibinfo{publisher}{Elsevier Ltd.}, \bibinfo{address}{Oxford},
  \bibinfo{year}{2011}), \bibinfo{edition}{5th editio} edn.

\bibitem{Abbey2013}
\bibinfo{author}{Abbey, B.}
\newblock \bibinfo{journal}{\bibinfo{title}{From grain boundaries to single
  defects: A review of coherent methods for materials imaging in the x-ray
  sciences}}.
\newblock {\emph{\JournalTitle{J. Miner. Met. Mater. Soc.}}}
  \textbf{\bibinfo{volume}{65}}, \bibinfo{pages}{1183–1201}
  (\bibinfo{year}{2013}).

\bibitem{Campbell2014}
\bibinfo{author}{Campbell, G.~H.}, \bibinfo{author}{McKeown, J.~T.} \&
  \bibinfo{author}{Santala, M.~K.}
\newblock \bibinfo{journal}{\bibinfo{title}{Time resolved electron microscopy
  for in situ experiments}}.
\newblock {\emph{\JournalTitle{Applied Physics Review}}}
  \textbf{\bibinfo{volume}{1}} (\bibinfo{year}{2014}).

\bibitem{yildirim2022snake}
\bibinfo{author}{Yildirim, C.} \emph{et~al.}
\newblock \bibinfo{journal}{\bibinfo{title}{4d microstructural evolution in a
  heavily deformed ferritic alloy: A new perspective in recrystallisation
  studies}}.
\newblock {\emph{\JournalTitle{Scripta Materialia}}}
  \textbf{\bibinfo{volume}{214}}, \bibinfo{pages}{114689}
  (\bibinfo{year}{2022}).

\bibitem{Jakobsen2019}
\bibinfo{author}{Jakobsen, A.} \emph{et~al.}
\newblock \bibinfo{journal}{\bibinfo{title}{Mapping of individual dislocations
  with dark-field x-ray microscopy}}.
\newblock {\emph{\JournalTitle{Journal of Applied Crystallography}}}
  \textbf{\bibinfo{volume}{52}}, \bibinfo{pages}{122--132}
  (\bibinfo{year}{2019}).

\bibitem{Simons2018}
\bibinfo{author}{Simons, H.} \emph{et~al.}
\newblock \bibinfo{journal}{\bibinfo{title}{Long-range symmetry breaking in
  embedded ferroelectrics}}.
\newblock {\emph{\JournalTitle{Nature materials}}}
  \textbf{\bibinfo{volume}{17}}, \bibinfo{pages}{814--819}
  (\bibinfo{year}{2018}).

\bibitem{Sangid2020}
\bibinfo{author}{Gustafson, S.} \emph{et~al.}
\newblock \bibinfo{journal}{\bibinfo{title}{Quantifying microscale drivers for
  fatigue failure via coupled synchrotron x-ray characterization and
  simulations}}.
\newblock {\emph{\JournalTitle{Nature communications}}}
  \textbf{\bibinfo{volume}{11}}, \bibinfo{pages}{1--10} (\bibinfo{year}{2020}).

\bibitem{DresselhausMarais2021}
\bibinfo{author}{Dresselhaus-Marais, L.~E.} \emph{et~al.}
\newblock \bibinfo{journal}{\bibinfo{title}{In-situ visualization of long-range
  defect interactions at the edge of melting}}.
\newblock {\emph{\JournalTitle{Science Advances}}}
  \textbf{\bibinfo{volume}{7}}, \bibinfo{pages}{eabe8311}
  (\bibinfo{year}{2021}).

\bibitem{Yabashi2017}
\bibinfo{author}{Yabashi, M.} \& \bibinfo{author}{Tanaka, H.}
\newblock \bibinfo{journal}{\bibinfo{title}{{The next ten years of X-ray
  science}}}.
\newblock {\emph{\JournalTitle{Nature Photonics}}}
  \textbf{\bibinfo{volume}{11}}, \bibinfo{pages}{12--14}
  (\bibinfo{year}{2017}).

\bibitem{Chapman2010}
\bibinfo{author}{Chapman, H.~N.} \emph{et~al.}
\newblock \bibinfo{journal}{\bibinfo{title}{Femtosecond diffractive imaging
  with a soft-x-ray free-electron laser}}.
\newblock {\emph{\JournalTitle{Nature Physics}}} \textbf{\bibinfo{volume}{2}},
  \bibinfo{pages}{839–843} (\bibinfo{year}{2006}).

\bibitem{Gorhover2016}
\bibinfo{author}{Gorkhover, T.} \emph{et~al.}
\newblock \bibinfo{journal}{\bibinfo{title}{Femtosecond and nanometre
  visualization of structural dynamics in superheated nanoparticles}}.
\newblock {\emph{\JournalTitle{Nature Photonics}}}
  \textbf{\bibinfo{volume}{10}}, \bibinfo{pages}{93–97}
  (\bibinfo{year}{2016}).

\bibitem{Robinson2009}
\bibinfo{author}{Robinson, I.} \& \bibinfo{author}{Harder, R.}
\newblock \bibinfo{journal}{\bibinfo{title}{Coherent x-ray diffraction imaging
  of strain at the nanoscale}}.
\newblock {\emph{\JournalTitle{Nature Materials}}}
  \textbf{\bibinfo{volume}{8}}, \bibinfo{pages}{291--298}
  (\bibinfo{year}{2009}).

\bibitem{Sakdinawat2010}
\bibinfo{author}{Sakdinawat, A.} \& \bibinfo{author}{Attwood, D.}
\newblock \bibinfo{journal}{\bibinfo{title}{Nanoscale x-ray imaging}}.
\newblock {\emph{\JournalTitle{Nature Photonics}}}
  \textbf{\bibinfo{volume}{4}}, \bibinfo{pages}{840–848}
  (\bibinfo{year}{2010}).

\bibitem{Kim2018nanoparticle}
\bibinfo{author}{Kim, D.} \emph{et~al.}
\newblock \bibinfo{journal}{\bibinfo{title}{Active site localization of methane
  oxidation on pt nanocrystals}}.
\newblock {\emph{\JournalTitle{Nature Communications}}}
  \textbf{\bibinfo{volume}{9}}, \bibinfo{pages}{3422} (\bibinfo{year}{2018}).

\bibitem{barbato2022phasex}
\bibinfo{author}{Barbato, F.}, \bibinfo{author}{Atzeni, S.},
  \bibinfo{author}{Batani, D.} \& \bibinfo{author}{Antonelli, L.}
\newblock \bibinfo{journal}{\bibinfo{title}{Phasex: an x-ray phase-contrast
  imaging simulation code for matter under extreme conditions}}.
\newblock {\emph{\JournalTitle{Optics Express}}} \textbf{\bibinfo{volume}{30}},
  \bibinfo{pages}{3388--3403} (\bibinfo{year}{2022}).

\bibitem{Poulsen2017}
\bibinfo{author}{Poulsen, H.~F.} \emph{et~al.}
\newblock \bibinfo{journal}{\bibinfo{title}{X-ray diffraction microscopy based
  on refractive optics}}.
\newblock {\emph{\JournalTitle{Journal of Applied Crystallography}}}
  \textbf{\bibinfo{volume}{50}}, \bibinfo{pages}{1441–1456}
  (\bibinfo{year}{2017}).

\bibitem{Poulsen2018}
\bibinfo{author}{Poulsen, H.} \emph{et~al.}
\newblock \bibinfo{journal}{\bibinfo{title}{Reciprocal space mapping and strain
  scanning using x-ray diffraction microscopy}}.
\newblock {\emph{\JournalTitle{Journal of Applied Crystallography}}}
  \textbf{\bibinfo{volume}{51}}, \bibinfo{pages}{1428--1436}
  (\bibinfo{year}{2018}).

\bibitem{Poulsen2021}
\bibinfo{author}{Poulsen, H.~F.}, \bibinfo{author}{Dresselhaus-Marais, L.~E.},
  \bibinfo{author}{Carlsen, M.~A.}, \bibinfo{author}{Winther, G.} \&
  \bibinfo{author}{Detlefs, C.}
\newblock \bibinfo{journal}{\bibinfo{title}{{Geometrical Optics Formalism to
  Model Contrast in Dark-Field X-ray Microscopy}}}.
\newblock {\emph{\JournalTitle{In Review, arXiv: 2007.09475}}}
  (\bibinfo{year}{2021}).

\bibitem{Holstad2022}
\bibinfo{author}{Holstad, T.~S.} \emph{et~al.}
\newblock \bibinfo{journal}{\bibinfo{title}{X-ray free-electron laser based
  dark-field x-ray microscopy: a simulation-based study}}.
\newblock {\emph{\JournalTitle{Journal of Applied Crystallography}}}
  \textbf{\bibinfo{volume}{55}} (\bibinfo{year}{2022}).

\bibitem{busing1967goniometer}
\bibinfo{author}{Busing, W.~R.} \& \bibinfo{author}{Levy, H.~A.}
\newblock \bibinfo{journal}{\bibinfo{title}{Angle calculations for 3-and
  4-circle x-ray and neutron diffractometers}}.
\newblock {\emph{\JournalTitle{Acta Crystallographica}}}
  \textbf{\bibinfo{volume}{22}}, \bibinfo{pages}{457--464}
  (\bibinfo{year}{1967}).

\bibitem{alonso2015x}
\bibinfo{author}{Alonso-Mori, R.} \emph{et~al.}
\newblock \bibinfo{journal}{\bibinfo{title}{The x-ray correlation spectroscopy
  instrument at the linac coherent light source}}.
\newblock {\emph{\JournalTitle{Journal of synchrotron radiation}}}
  \textbf{\bibinfo{volume}{22}}, \bibinfo{pages}{508--513}
  (\bibinfo{year}{2015}).

\bibitem{feng2011single}
\bibinfo{author}{Feng, Y.} \emph{et~al.}
\newblock \bibinfo{title}{A single-shot intensity-position monitor for hard
  x-ray fel sources}.
\newblock In \emph{\bibinfo{booktitle}{X-ray Lasers and Coherent X-ray Sources:
  Development and Applications IX}}, vol. \bibinfo{volume}{8140},
  \bibinfo{pages}{163--168} (\bibinfo{organization}{SPIE},
  \bibinfo{year}{2011}).

\bibitem{Ishikawa2005}
\bibinfo{author}{Ishikawa, T.}, \bibinfo{author}{Tamasaku, K.} \&
  \bibinfo{author}{Yabashi, M.}
\newblock \bibinfo{journal}{\bibinfo{title}{High-resolution x-ray
  monochromators}}.
\newblock {\emph{\JournalTitle{Nuclear Instruments and Methods in Physics
  Research Section A: Accelerators, Spectrometers, Detectors and Associated
  Equipment}}} \textbf{\bibinfo{volume}{547}}, \bibinfo{pages}{42--49}
  (\bibinfo{year}{2005}).

\bibitem{koch1998x}
\bibinfo{author}{Koch, A.}, \bibinfo{author}{Raven, C.},
  \bibinfo{author}{Spanne, P.} \& \bibinfo{author}{Snigirev, A.}
\newblock \bibinfo{journal}{\bibinfo{title}{X-ray imaging with submicrometer
  resolution employing transparent luminescent screens}}.
\newblock {\emph{\JournalTitle{Journal of the Optical Society of America A}}}
  \textbf{\bibinfo{volume}{15}}, \bibinfo{pages}{1940--1951}
  (\bibinfo{year}{1998}).

\bibitem{Breckling2022}
\bibinfo{author}{Breckling, S.} \emph{et~al.}
\newblock \bibinfo{journal}{\bibinfo{title}{An automated approach to the
  alignment of compound refractive lenses}}.
\newblock {\emph{\JournalTitle{Journal of Synchrotron Radiation}}}
  \textbf{\bibinfo{volume}{29}} (\bibinfo{year}{2022}).

\bibitem{kudo2022x}
\bibinfo{author}{Kudo, T.} \emph{et~al.}
\newblock \bibinfo{journal}{\bibinfo{title}{An x-ray beam profile monitoring
  system at a beamline front-end combining a single-crystal diamond film and
  energy discrimination using droplet analysis}}.
\newblock {\emph{\JournalTitle{Journal of Synchrotron Radiation}}}
  \textbf{\bibinfo{volume}{29}} (\bibinfo{year}{2022}).

\bibitem{vandriel2015timingtool}
\bibinfo{author}{van Driel, T.~B.} \emph{et~al.}
\newblock \bibinfo{journal}{\bibinfo{title}{Disentangling detector data in xfel
  studies of temporally resolved solution state chemistry}}.
\newblock {\emph{\JournalTitle{Faraday discussions}}}
  \textbf{\bibinfo{volume}{177}}, \bibinfo{pages}{443--465}
  (\bibinfo{year}{2015}).

\bibitem{geloni2011selfseeding}
\bibinfo{author}{Geloni, G.}, \bibinfo{author}{Kocharyan, V.} \&
  \bibinfo{author}{Saldin, E.}
\newblock \bibinfo{journal}{\bibinfo{title}{A novel self-seeding scheme for
  hard x-ray fels}}.
\newblock {\emph{\JournalTitle{Journal of Modern Optics}}}
  \textbf{\bibinfo{volume}{58}}, \bibinfo{pages}{1391--1403}
  (\bibinfo{year}{2011}).

\bibitem{kutsal2019esrf}
\bibinfo{author}{Kutsal, M.} \emph{et~al.}
\newblock \bibinfo{title}{The esrf dark-field x-ray microscope at id06}.
\newblock In \emph{\bibinfo{booktitle}{IOP conference series: materials science
  and engineering}}, vol. \bibinfo{volume}{580}, \bibinfo{pages}{012007}
  (\bibinfo{organization}{IOP Publishing}, \bibinfo{year}{2019}).

\bibitem{saccocio1965borrmann}
\bibinfo{author}{Saccocio, E.~J.} \& \bibinfo{author}{Zajac, A.}
\newblock \bibinfo{journal}{\bibinfo{title}{Simultaneous diffraction of x rays
  and the borrmann effect}}.
\newblock {\emph{\JournalTitle{Physical Review}}}
  \textbf{\bibinfo{volume}{139}}, \bibinfo{pages}{A255} (\bibinfo{year}{1965}).

\bibitem{Holstad2022b}
\bibinfo{author}{Holstad, T.~S.} \emph{et~al.}
\newblock \bibinfo{journal}{\bibinfo{title}{Real-time imaging of acoustic waves
  in bulk materials with x-ray microscopy}}.
\newblock {\emph{\JournalTitle{In Preparation}}}  (\bibinfo{year}{2022}).

\bibitem{Simons2015}
\bibinfo{author}{Simons, H.} \emph{et~al.}
\newblock \bibinfo{journal}{\bibinfo{title}{Dark-field x-ray microscopy for
  multiscale structural characterization}}.
\newblock {\emph{\JournalTitle{Nature Communications}}}
  \textbf{\bibinfo{volume}{6}}, \bibinfo{pages}{6098} (\bibinfo{year}{2015}).

\bibitem{Ferrer2022darfix}
\bibinfo{author}{Ferrer, J.~G.} \emph{et~al.}
\newblock \bibinfo{journal}{\bibinfo{title}{darfix: Data analysis for
  dark-field x-ray microscopy}}.
\newblock {\emph{\JournalTitle{arXiv preprint arXiv:2205.05494}}}
  (\bibinfo{year}{2022}).

\bibitem{Poulsen2004}
\bibinfo{author}{Poulsen, H.~F.}
\newblock \emph{\bibinfo{title}{Three-dimensional X-ray diffraction microscopy:
  mapping polycrystals and their dynamics}}, vol. \bibinfo{volume}{205}
  (\bibinfo{publisher}{Springer Science \& Business Media},
  \bibinfo{year}{2004}).

\bibitem{Carlsen2022}
\bibinfo{author}{Carlsen, M.}, \bibinfo{author}{Detlefs, C.},
  \bibinfo{author}{Yildirim, C.}, \bibinfo{author}{R{\ae}der, T.} \&
  \bibinfo{author}{Simons, H.}
\newblock \bibinfo{journal}{\bibinfo{title}{Simulating dark-field x-ray
  microscopy images with wave front propagation techniques}}.
\newblock {\emph{\JournalTitle{arXiv preprint arXiv:2201.07549}}}
  (\bibinfo{year}{2022}).

\bibitem{brennan2022analytical}
\bibinfo{author}{Brennan, M.~C.}, \bibinfo{author}{Howard, M.},
  \bibinfo{author}{Marzouk, Y.} \& \bibinfo{author}{Dresselhaus-Marais, L.~E.}
\newblock \bibinfo{journal}{\bibinfo{title}{Analytical methods for
  superresolution dislocation identification in dark-field x-ray microscopy}}.
\newblock {\emph{\JournalTitle{Journal of Materials Science}}}
  \textbf{\bibinfo{volume}{57}}, \bibinfo{pages}{14890--14904}
  (\bibinfo{year}{2022}).

\bibitem{Legge1987pedestal}
\bibinfo{author}{Legge, G.~E.}, \bibinfo{author}{Kersten, D.} \&
  \bibinfo{author}{Burgess, A.~E.}
\newblock \bibinfo{journal}{\bibinfo{title}{Contrast discrimination in noise}}.
\newblock {\emph{\JournalTitle{JOSA A}}} \textbf{\bibinfo{volume}{4}},
  \bibinfo{pages}{391--404} (\bibinfo{year}{1987}).

\bibitem{Li2011}
\bibinfo{author}{Li, H.}, \bibinfo{author}{Fu, M.}, \bibinfo{author}{Lu, J.} \&
  \bibinfo{author}{Yang, H.}
\newblock \bibinfo{journal}{\bibinfo{title}{Ductile fracture: experiments and
  computations}}.
\newblock {\emph{\JournalTitle{International Journal of Plasticity}}}
  \textbf{\bibinfo{volume}{27}}, \bibinfo{pages}{147--180}
  (\bibinfo{year}{2011}).

\bibitem{martin2012femtosecond}
\bibinfo{author}{Martin, A.~V.} \emph{et~al.}
\newblock \bibinfo{journal}{\bibinfo{title}{Femtosecond dark-field imaging with
  an x-ray free electron laser}}.
\newblock {\emph{\JournalTitle{Optics express}}} \textbf{\bibinfo{volume}{20}},
  \bibinfo{pages}{13501--13512} (\bibinfo{year}{2012}).

\bibitem{schropp2013full}
\bibinfo{author}{Schropp, A.} \emph{et~al.}
\newblock \bibinfo{journal}{\bibinfo{title}{Full spatial characterization of a
  nanofocused x-ray free-electron laser beam by ptychographic imaging}}.
\newblock {\emph{\JournalTitle{Scientific reports}}}
  \textbf{\bibinfo{volume}{3}}, \bibinfo{pages}{1633} (\bibinfo{year}{2013}).

\bibitem{daurer2021ptychographic}
\bibinfo{author}{Daurer, B.~J.} \emph{et~al.}
\newblock \bibinfo{journal}{\bibinfo{title}{Ptychographic wavefront
  characterization for single-particle imaging at x-ray lasers}}.
\newblock {\emph{\JournalTitle{Optica}}} \textbf{\bibinfo{volume}{8}},
  \bibinfo{pages}{551--562} (\bibinfo{year}{2021}).

\bibitem{ulvestad2017bragg}
\bibinfo{author}{Ulvestad, A.} \emph{et~al.}
\newblock \bibinfo{journal}{\bibinfo{title}{Bragg coherent diffractive imaging
  of zinc oxide acoustic phonons at picosecond timescales}}.
\newblock {\emph{\JournalTitle{Scientific reports}}}
  \textbf{\bibinfo{volume}{7}}, \bibinfo{pages}{9823} (\bibinfo{year}{2017}).

\bibitem{Clark2015}
\bibinfo{author}{Clark, J.~N.} \emph{et~al.}
\newblock \bibinfo{journal}{\bibinfo{title}{Three-dimensional imaging of
  dislocation propagation during crystal growth and dissolution}}.
\newblock {\emph{\JournalTitle{Nature Materials}}}
  \textbf{\bibinfo{volume}{14}}, \bibinfo{pages}{780–784}
  (\bibinfo{year}{2015}).

\bibitem{kang2020bcdi_catalysis}
\bibinfo{author}{Kang, J.} \emph{et~al.}
\newblock \bibinfo{journal}{\bibinfo{title}{Time-resolved in situ visualization
  of the structural response of zeolites during catalysis}}.
\newblock {\emph{\JournalTitle{Nature communications}}}
  \textbf{\bibinfo{volume}{11}}, \bibinfo{pages}{5901} (\bibinfo{year}{2020}).

\bibitem{carlsen2022fourier}
\bibinfo{author}{Carlsen, M.} \emph{et~al.}
\newblock \bibinfo{journal}{\bibinfo{title}{Fourier ptychographic dark field
  x-ray microscopy}}.
\newblock {\emph{\JournalTitle{Optics Express}}} \textbf{\bibinfo{volume}{30}},
  \bibinfo{pages}{2949--2962} (\bibinfo{year}{2022}).

\end{thebibliography}

\section*{Acknowledgements}

This work was performed in part under the auspices of the US Department of Energy by Lawrence Livermore National Laboratory under contract DE-AC52-07NA27344. We also acknowledge the Lawrence Fellowship, which funded developmental contributions from LEDM in this work.
TSH acknowledges financial support
from Villum FONDEN (grant no. 00028346), HFP from the Danish Agency for Science and Higher Education (grant number 8144-00002B) and from the European Research Council (Advanced grant no 885022 and Starting grant no
804665).
This work was authored in part by Mission Support and Test Services, LLC, under Contract No. DE-NA0003624 with the U.S. Department of Energy, the Office of Defense Programs, and supported by the Site-Directed Research and Development Program. DOE/NV/03624-{}-1478.
D.N. was also supported by the National Research Foundation of Korea (NRF-2021R1F1A1051444 and NRF-2022M3H4A1A04074153). HK, SK and SK were funded by the National Research Foundation of Korea (NRF- 2021R1A3B1077076).
RSM acknowledges support from EPSRC First Grant EP/P024513/1 and ERC Consolidator Grant TRIREME.
TMR acknowledges funding from the European Union’s Horizon 2020 research and innovation programme under the Marie Skłodowska-Curie grant agreement No 899987.

\section*{Author contributions statement}

LDM wrote the manuscript with significant contributions from TH, HFP, MMN, KH and TR, then edits and correspondence from all co-authors.
LDM designed the experimental configurations, refined experiments, led experiments and analysis. 
Spontaneous acquisition experiments were conceived of at LLNL, while the pump-probe experiments were conceived of at DTU.
EF developed the 3D models and engineering designs for the U-HXM instrument and helped to construct them at LCLS and PAL-XFEL.
BK and TMR performed data analysis and image acquisitions at the LCLS, supported by TSH, and EBK.
Further post-processing data analysis were also performed by LEDM, KH, and ZS. 
XFEL experiments were carried out and supported by LEDM, BK, TSH, MS, DN, SB, PKC, JHE, EF, EG, LG, AG, KH, MH, KK, SK, SK, SK, SK, HK, EBK, SK, H-JL, CL, RSM, BN, NO, RPP, HFP, MMN, TMR, AMS, TS, HS, TvD, BW, WY, CY.
DN, SK, SN, MS, TvD, and MC supported the experiments as beamline scientists at the PAL-XFEL and LCLS, respectively.
FS evaluated and selected the diamond samples for all experiments presented in this work.
SB, AG, and MH supported the analysis of data collected at PAL-XFEL and developed optimization methodologies in support of the optics alignment in this work.

\section*{Data Availability}

The datasets used and/or analysed during the current study available from the corresponding author on reasonable request.

\section*{Additional information}

The authors of this work have no competing interests in this work. 

\end{document}